\documentclass[10pt]{iopart}
\usepackage{graphicx}
\usepackage{harvard}

\begin{document}

\jl{21}

\topical{Andreev bound states in high-$T_c$ superconducting junctions}

\author{T L\"ofwander, V S Shumeiko and G Wendin}

\address{Department of Microelectronics and Nanoscience,\\
School of Physics and Engineering Physics,\\
Chalmers University of Technology and G\"oteborg University,\\
S-412 96 G\"oteborg,
Sweden}

\begin{abstract}
  The formation of bound states at surfaces of materials with an
  energy gap in the bulk electron spectrum is a well known physical
  phenomenon. At superconductor surfaces, quasiparticles with energies
  inside the superconducting gap $\Delta$ may be trapped in bound
  states in quantum wells, formed by total reflection against the
  vacuum and total Andreev reflection against the superconductor.
  Since an electron reflects as a hole and sends a Cooper pair into
  the superconductor, the surface states give rise to resonant
  transport of quasiparticle and Cooper pair currents, and may be
  observed in tunneling spectra. In superconducting junctions, these
  surface states may hybridize and form bound Andreev states, trapped
  between the superconducting electrodes. In $d$-wave superconductors,
  the order parameter changes sign under $90^o$ rotation and, as a
  consequence, Andreev reflection may lead to the formation of zero
  energy quasiparticle bound states, midgap states (MGS). The
  formation of MGS is a robust feature of d-wave superconductivity and
  provides a unified framework for many important effects which will
  be reviewed: large Josephson current, low-temperature anomaly of the
  critical Josephson current, $\pi$-junction behavior, $0\rightarrow
  \pi$ junction crossover with temperature, zero-bias conductance
  peaks, paramagnetic currents, time reversal symmetry breaking,
  spontaneous interface currents, and resonance features in subgap
  currents. Taken together these effects, when observed in
  experiments, provide proof for $d$-wave superconductivity in the
  cuprates.
\end{abstract}

\submitted


\maketitle

\citationstyle{dcu}
\citationmode{abbr}

\pagestyle{empty}
\tableofcontents

\section{Introduction}
\pagestyle{headings}

Ever since the discovery of high-$T_c$ superconductivity \cite{BM}
spin-fluctuation mediated d-wave singlet pairing has been a major
candidate for superconductive pairing in the cuprates
\cite{Dagotto,Scalapino}.  Already from the start it was realized that
the layered cuprates are low dimensional systems with
antiferromagnetic undoped parent compounds, with strong electron
correlations and with highly unusual normal electron properties.
Nevertheless, phonon-mediated strong-coupling s-wave singlet pairing
could not be ruled out because of the experimental difficulty to
identify essential d-wave signatures which cannot also be explained
within an s-wave picture. Looking back, it seems that the essential
quantitative body of facts, not directly addressing the qualitative
question of symmetry, was known already by 1990.  In particular, one
early experimental result may be regarded as a clear qualitative
signature of d-wave pairing: the so called zero-bias conductance peak
(ZBCP) observed in the I-V characteristics of electron tunneling into
thin films of YBaCaCuO already by \citeasnoun{Geerk}. However, the
d-wave connection was not realized at the time; it was later proposed
by \citeasnoun{Hu} and \citeasnoun{Kash95} and then experimentally
verified by \citeasnoun{Cov} in a study of the magnetic field
dependence of the splitting of the ZBCP.

A large effort has gone into finding the most significant evidence for
d-wave pairing symmetry of the order parameter in the high-$T_c$
cuprates [see e.g. the review papers by \citeasnoun{Dagotto},
\citeasnoun{Scalapino} and \citeasnoun{Annett} and references
therein]. The problem is nontrivial because it is necessary to
distinguish $d$-wave symmetry from anisotropic $s$-wave symmetry. A
$d$-wave order parameter has nodes at four mutually perpendicular
directions of the electron momentum. This provides gapless
superconductivity in the direction of the nodes, with normal
excitations existing at all energies, having far-reaching consequences
for the materials properties of the superconducting state, e.g. linear
heat capacity at low temperature, large subgap tunnel current, smeared
gap structures and high sensitivity to non-magnetic impurities. These
properties have indeed been found in experiments, from the beginning.
Moreover, anisotropic gap structures with deep minima in the node
directions have been established in angle resolved photoemission
experiments. The problem is, however, that these experimental facts
are also consistent with an anisotropic $s$-wave order parameter [see
the review by \citeasnoun{Annett} for an analysis of experimental
results].

To make an "experimentum crucius" one has to establish that the order
parameter {\em changes sign} when the momentum rotates by 90 degrees.
Such experiments have earlier been suggested by \citeasnoun{Larkin} in
connection with heavy fermion superconductivity, and first realized in
tunneling experiments on high-$T_c$ superconductors by
\citeasnoun{Wollman1}. The idea behind these experiments is based on the
phase sensitive properties of Josephson junctions involving $d$-wave
and $s$-wave, or two misoriented $d$-wave, superconducting electrodes.
When the junction electrodes have different signs of the order
parameter, this is equivalent to adding a $\pi$-shift to the
superconducting phase difference across the junction, and therefore
the stable state of the junction corresponds to a superconducting
phase difference equal to $\phi =\pi$ ($\pi$-junctions), instead of
$\phi = 0$ like in conventional junctions.

The conceptually simplest way to create a $\pi$-junction is to attach
a superconducting loop to two adjacent faces at right angles of a
high-$T_c$ single crystal, creating a SQUID circuit with one
$\pi$-junction \cite{Mathai,Wollman2,Harl,Mannhart_APL00}. Using
tri-crystal substrates one can create a superconducting ring
consisting of an odd number of $\pi$-junctions which is able to trap
half-integer flux quanta. This was studied by \citename{tricrystal1}
\citeyear{tricrystal1,tricrystal4,tricrystal5} and Kirtley et al.
\citeyear{tricrystal2,tricrystal3} who used a scanning SQUID to detect
spontaneous half-integer flux quanta in the $\pi$-junction loop but no
flux quanta in rings on bi-crystal interfaces; for an overview of the
subject, see the recent review article by \citeasnoun{RevTK}.  This is
in principle the definitive experiment, but it is desirable that it
can be repeated by other groups.

There is however another type of experiments which may provide direct
evidence for $d$-wave symmetry of the order parameter.  These
experiments exploit the property of the $d$-wave superconductors to
form superconducting bound surface states at the Fermi energy, so
called midgap states, MGS \cite{Hu}. MGS is a direct consequence of
the angle-dependent sign change of the order parameter, and it cannot
exist in $s$-wave superconductors. Transport measurements which detect
the MGS resonance at the Fermi energy for particular orientations of
the plane of the junction with respect to the crystallographic axes
will provide crucial evidence for $d$-wave superconductivity.

The problem of MGS is part of a general problem of superconducting
surface and interface states and their relation to the Andreev states,
which determine the transport properties of superconducting junctions.
This is the topic of the present review.

\section{Superconducting surface states}

As is well known, at semiconductor and metal surfaces quasiparticle
surface states may exist in the band gaps of the bulk material
\citeaffixed{AshcroftMermin}{see e.g.}. The situation is similar at a
superconductor surface: the important difference is, however, that on
a superconductor surface, electrons are reflected as holes and holes
are reflected as electrons - Andreev reflection \cite{Andreev} - while
on a normal surface electrons are reflected in the normal way as
electrons.

Superconducting quasiparticle surface states were first described by
\citeasnoun{deG_StJ} for the INS quantum well. More recently it was
found that the superconducting surface states do not need a normal
region at the surface and they exist at an arbitrary interface between
an s-wave superconductor and an insulator \cite{Giant}.  The energy of
the surface states lies within the energy gap of the superconductor
and it depends on the properties of the surface. In d-wave
superconductors, due to the specific symmetry of the order parameter,
these levels may occur exactly at the Fermi energy \cite{Hu,Hu_PRB98}.

In superconducting junctions, the overlap of the wave functions of the
surface states builds up stationary states at certain energies, the
Andreev bound states. These states are thus states of the entire
junction, similar to bonding and antibonding states in a molecule or
in a system of two closely separated metal surfaces. The name of these
bound states are associated with Andreev, who first studied them in
perfect SNS junctions \cite{Andreev_SNS}. Due to their ability to
carry current, the Andreev bound states play an essential role in dc
Josephson transport \cite{Kulik,Bardeen-Johnson}. Moreover, the
surface states in superconductor-normal metal and
superconductor-superconductor junctions introduce resonances in the
current transport under applied voltage bias.

\subsection{Andreev reflection}

An important aspect of the superconducting surface states is that they
are built up from a combination of electron and hole wave functions.
This mixing is induced by the specific property of the superconductor
interface known as {\em Andreev reflection}.

The problem was originally raised by the observation of an anomalously
large thermal resistance of superconductors in the intermediate state
\cite{mendelssohn,zavaritsky}. It was suggested that the enhancement
of the thermal resistance was due to some strong additional
quasiparticle scattering at the NS interfaces \cite{hulm,strassler}.
However, no microscopic mechanism for such a reflection was proposed
because the spatial variation of the order parameter in the
intermediate state is smooth on the scale of the Fermi wave length,
and it was not clear how such a "soft" potential could provide
efficient electron reflection.  The problem was solved by Andreev who
showed that the reflection at NS interfaces is due to electron-hole
conversion \cite{Andreev}.

Let us consider a contact between a superconductor and a normal metal,
and assume for simplicity that the normal electron properties of both
electrodes are identical. This means that normal electrons will not be
reflected by the contact - perfect transmission in the normal state.
However, electrons in the normal electrode which approach the
interface with an energy E inside the superconducting gap,
$|E|<\Delta$, cannot penetrate into the superconducting electrode. At
the same time, there is no normal scattering process at the interface
which is able to reverse the electron momentum. The problem is solved
by converting the incoming electron into a {\em reflected hole} moving
with the {\em same momentum} in the opposite direction: since the hole
energy is given by $E^h=E_F-p^2/2m$, the group velocity is
$v^h=dE^h/dp\approx -v_F$.  Such a conversion, Andreev reflection,
occurs due to electron-hole correlations within the superconducting
electrode.

A straightforward quantum mechanical analysis (see the appendix) shows
that Andreev reflection is accompanied by a phase shift
$-\gamma(E)\mp\chi$, where the upper sign corresponds to the
conversion of an electron into a hole, and the lower sign to the
inverse process. In this expression, $\gamma(E)=\arccos
(E/|\Delta|)$ is the Andreev reflection phase shift and $\chi$ is the
phase of the superconducting order parameter. The wave functions of
the electrons and holes penetrate into the superconductor over a
length of the order of the superconducting coherence length,
$\xi_0=\hbar v_F/\Delta$, and therefore the electron-hole conversion
occurs over this distance.  Thus, the phase shift during Andreev
reflection is sensitive to the profile $\Delta(x)$ of the
superconducting order parameter over a distance of the order of a
coherence length near the interface.  Calculation of this profile
involves self-consistent treatment of the order parameter: however,
the properties of the surface states can be qualitatively understood
by replacing the exact form of $\Delta(x)$ by a step function
$\Delta(x)=\Delta\Theta(x)$.

\subsection{Surface states}\label{surf_section}

\begin{figure}
\centerline{\includegraphics[width=8cm]{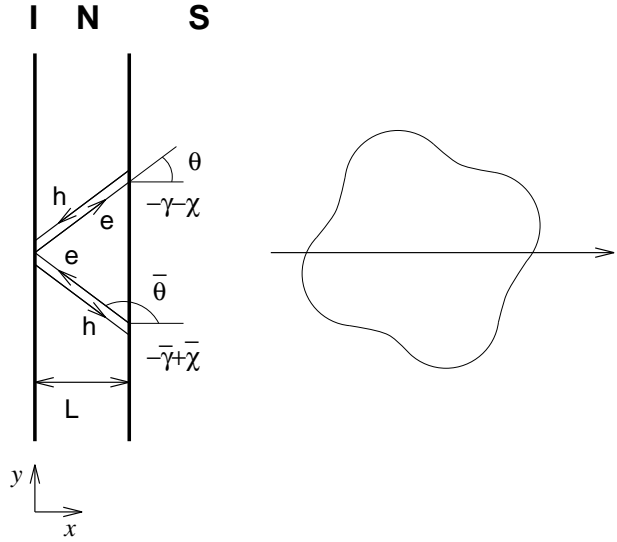}}
\caption{Quasiclassical path giving the bound states at a specular
  surface of a superconductor with anisotropic s-wave order parameter
  $\Delta(\theta)$ (schematically drawn in the right part of the
  figure). N is a normal region of size $L\sim\xi_0$ to model the gap
  suppression near the surface. The quasiparticles are trapped in the
  surface region because of normal reflection at the vacuum/metal (IN)
  surface and subgap Andreev reflection at the NS surface. The angle
  $\theta$ gives the propagation direction relative to the surface
  normal. During Andreev reflection, an electron (e) is converted into
  a hole (h) with the phase gain $-\gamma-\chi$.  According to the
  Bohr-Sommerfeld quantization condition, the total phase accumulated
  during one cycle is equal to $2\pi n$.}
\label{surface_s_fig}
\end{figure}

Let us now consider the formation of superconducting bound states in
an INS quantum well with a specular vacuum/metal (IN) interface
surface. We will first consider a two-dimensional s-wave
superconductor with anisotropic order parameter which has a perfectly
transparent normal metal/superconductor (NS) interface with a normal
layer of width $L$, as shown in \fref{surface_s_fig}.  According
to quasiclassical arguments, the bound state corresponds to a closed
quasiparticle trajectory, and the bound state energy is given by the
Bohr-Sommerfeld quantization condition which requires the total phase
accumulated during one cycle to be equal to an integer number of
$2\pi$. The quasiparticle trajectory in \fref{surface_s_fig}
(corresponding to electron motion in the $+k_y$-direction) consists of
an electron segment which includes a single reflection at the IN
interface, and of a hole segment which backwards retraces the electron
trajectory. The total accumulated phase consists of two parts: (i) a
contribution from the Andreev reflections: $-\gamma - \chi $ in the
$\theta$-direction and $-\bar\gamma + \bar\chi $ in the
$\bar\theta$-direction ($\theta$ is the angle between the electron
trajectory and the surface normal and $\bar\theta=\pi-\theta$ is the
angle after specular normal reflection at the surface), and (ii) a
contribution from the phase $\beta$ accumulated during propagation
through the normal region, $\beta = 2L(k^e-k^h) +\beta_0$, where the
first term corresponds to the ballistic motion and the second to the
reflection at the NI interface.  The electron/hole wave vectors are
$\hbar k^{e,h}= \sqrt{2m(E_F\cos^2\theta\pm E)}$. The
Bohr-Sommerfeld quantization condition can then be written down as
\begin{equation}\label{BS1}
-(\gamma +\bar\gamma) \mp \delta\chi + \beta(E) = 2n\pi,
\end{equation}
where $\delta\chi=\chi-\bar\chi$, and where $\mp \delta\chi$
corresponds to quasiparticle trajectories in the $\pm k_y$ directions.

In the case of a uniform and isotropic order parameter ($\gamma
=\bar\gamma; \; \chi =\bar\chi$), equation~\eref{BS1} reduces to
\begin{equation}\label{BS2}
-2\gamma + \beta = 2n\pi
\end{equation}
and leads to the spectral equation
\begin{equation}\label{surfstate_s}
E= \pm\Delta\cos{\beta(E)\over2}.
\end{equation}
According to this equation, surface states always exist, even at a
simple IS interface ($L=0$) due to the normal reflection phase shift
$\beta_0\neq 0$ in the superconducting electrode \cite{Giant}.
However, in the presence of an INS well, $L\neq 0$, the ballistic
contribution to $\beta$, $2L(k^e-k^h)\approx 4LE/\hbar v_F\cos\theta$,
usually dominates, and the dispersion equation \eref{surfstate_s}
takes the form found by \citeasnoun{deG_StJ},
\begin{equation}\label{ss_ballistic}
{E\over\Delta} = \pm \cos{{E\over\Delta}{2L\over{\xi_0\cos\theta}}}.
\end{equation}

\begin{figure}
\centerline{\includegraphics[width=8cm]{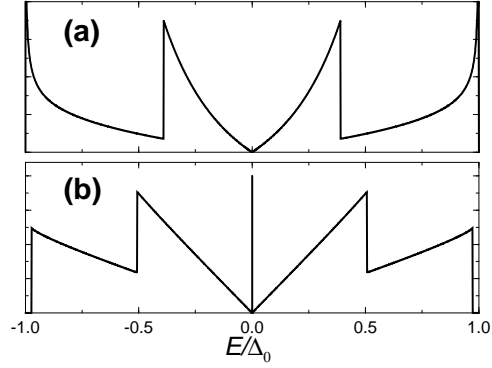}}
\caption{(a) Density of surface states at an $s$-wave superconductor
  surface with a normal metal overlayer of thickness $1.5\xi_0$. The
  peak energies are the bound states for trajectory angle $\theta=0$
  and the tails towards lower energies are due to the angle
  integration; for increasing angles $\theta$ the energy of the
  surface states are continuously shifted down relative to the
  $\theta=0$ state. (b) The same for a gap of $d$-wave symmetry. The
  crystal is rotated by $45^o$ relative to the surface normal so that
  midgap states are formed (the delta peak at zero
  energy).}\label{surfDOS_fig}
\end{figure}

Generally, the electron-hole dephasing factor $\beta(E)$ is an odd
function of the energy and turns to zero at $E=0$. This is a result of
the fact that the difference between electrons and holes vanishes at
the Fermi surface.  Due to this property there is no $E=0$ solution of
equation~\eref{surfstate_s}, and the density of surface states in
s-wave superconductors is zero at the Fermi level; see
\fref{surfDOS_fig}(a). Moreover, due to the presence of time reversal
symmetry in the problem, the surface levels always appear in pairs
symmetrically positioned around the Fermi level.

The number of surface states for each quasiparticle trajectory is
determined by the length of the quasiparticle path, which is given by
the normal layer thickness and the propagation angle $\theta$: for
$L<\pi\xi_0$, equation~\eref{ss_ballistic} implies that for normal incidence
($\theta=0$) there is a single pair of surface states close to the
energy gap edge. A new pair of surface states appear at the gap edge
when $1/\cos\theta=\pi\xi_0/L$. Increasing the trajectory angle
further pushes these states down into the gap, and a new pair of gap
edge states appears every time $1/\cos\theta$ is increased by
$\pi\xi_0/L$.  Integration over the angle $\theta$ yields the surface
density of states plotted in \fref{surfDOS_fig}(a), cf.  reference
\cite{deG_StJ}.

In anisotropic s-wave superconductors, different gap values before
($|\Delta|$) and after ($|\bar\Delta|$) normal reflection at the
surface ($\gamma \neq\bar\gamma$) will modify the phase shifts of the
Andreev reflections and change the energy of the surface state.
Therefore the surface states depend on how the surface is oriented
relative to the crystal axes. The angle integration will then to a
large extent wash out the peaks in the density of states (DOS) found
in \fref{surfDOS_fig}(a). Roughness of the interface will lead to
further suppression of the peaks in the DOS and to the opening of a
gap in the spectrum \cite{Melsen}. In this review we will however not
discuss rough interfaces in any detail. It is also worth mentioning
that for anisotropic superconductors, the total DOS within the energy
interval $\Delta_{min}<E<\Delta_{max}$, contains, in addition to the
bound surface states, contributions from the continuum states.

The surface states are two-fold degenerate, the two states
corresponding to different signs of the electron momentum ($\pm k_y$)
parallel to the surface. In the presence of a magnetic field parallel
to the surface, this degeneracy is removed due to the Meissner
current: the spectrum calculated above holds in the frame of reference
of the condensate. In the laboratory reference frame the levels are
shifted by the Doppler term $p_sv_F\sin\theta$, $E\rightarrow E+
p_sv_F\sin\theta$, where $p_s$ is the superfluid momentum of the
Meissner current. This shift has opposite signs for electrons moving
along and against the Meissner current.

\subsection{$d$-wave surface states}

\begin{figure}
\centerline{\includegraphics[width=8cm]{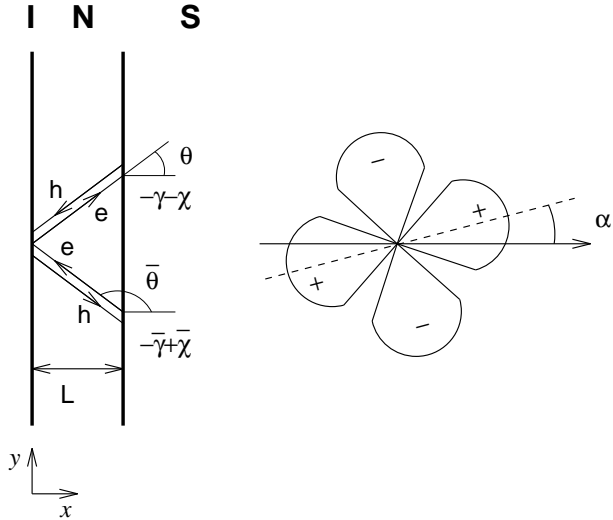}}
\caption{Quasiclassical path giving the bound states at a specular
  surface of a two-dimensional $d$-wave superconductor with order
  parameter $\Delta(\theta)=\Delta_0\cos[2(\theta-\alpha)]$ where
  $\theta$ is the propagation angle, compare Fig.~\ref{surface_s_fig},
  and $\alpha$ is the orientation angle of the order parameter
  relative to the surface normal. N is a normal region of size
  $L\sim\xi_0$ to model the gap suppression near the surface.
  }\label{surface_d_fig}
\end{figure}

In the d-wave superconductors, the order parameter is angle dependent,
\begin{equation}
\Delta(\theta)=\Delta_0\cos[2(\theta-\alpha)],
\end{equation}
where $\alpha$ describes the orientation of the order parameter
relative to the surface normal; see \fref{surface_d_fig}.
Consequently, the gap (order parameter) is different before ($\Delta$)
and after ($\bar\Delta$) normal reflection. As long as $\Delta$ and
$\bar\Delta$ have the same sign, i.e. the same phase, $\chi=\bar\chi$,
the situation is not qualitatively different from the anisotropic
s-wave case. However, the situation dramatically changes if $\Delta$
and $\bar\Delta$ have opposite signs, as illustrated in
\fref{surface_d_fig}: in this case, the phases $\chi$ and
$\bar\chi=\chi+\pi$ of these order parameters differ by
$\delta\chi=\chi-\bar\chi= -\pi$, which leads to a modification of
equation~\eref{BS2},
\begin{equation}\label{BS_dd}
-(\gamma+\bar\gamma) +\pi+ \beta(E) = 2n\pi.
\end{equation}
This equation always has the root $E=0$ since $\gamma(E=0)=
\bar{\gamma}(E=0)=\pi/2$ and $\beta(E=0)=0$. This means that for the
corresponding angle of incidence, a surface state will appear at
zero-energy: this is the so-called midgap state, MGS \cite{Hu}.
Clearly, this MGS is degenerate with respect to $\pm k_y$.  It is easy
to see that for the $\alpha=\pm \pi/4$ orientations, an MGS exists for
every quasiparticle trajectory, $-\pi/2<\theta<\pi/2$, while for the
$\alpha=0, \pi/2$ orientations there are no MGS at all. For arbitrary
$\alpha$, MGS exist within the window
$\pi/2-\alpha<|\theta|<\pi/2+\alpha$. The formation of MGS is a robust
property of d-wave superconductors: it requires only different signs
of $\Delta$ and $\bar\Delta$ and does not depend on the possible
difference in their absolute values. Moreover, the formation of MGS is
not sensitive to the spatial variation of $\Delta(x)$ (see
\ref{App_MGS}). This insensitivity is an important property, because
in d-wave superconductors the gap is suppressed by the surface for
most orientations of the crystallographic axes.

There is work in the literature addressing the questions of how
surface roughness \cite{Golubov}, impurities \cite{Poenicke}, and
nanofaceting of the surfaces/interfaces \cite{FRS,TTYK} affect the
MGS.  However, in this review we will focus on clean ballistic
structures with specularly reflecting interfaces, and we refer the
reader to the original literature (e.g. the abovementioned papers and
references therein).

The major effect of taking gap suppression into account will be the
presence of additional bound states in the vicinity of the gap edge.
Fully self-consistent calculations have been performed by e.g.
\cite{Nagato_prb95,BPRS1,BPRS2,MS2,MS3,MS4,BSB}. Qualitatively, the
effect of the surface levels at finite energy can be understood by
considering a rectangular SNI well of size $L\sim\xi_0$. For the sake
of illustration, let us consider the $\alpha=\pi/4$ orientation for
which an MGS exists for every trajectory angle $\theta$ (see
\fref{surface_d_fig}). We then have $|\Delta|=|\bar\Delta|$, and
equation~\eref{BS_dd} reduces to
\begin{equation}\label{ss_d_ballistic}
\frac{E}{|\Delta|} =\pm\sin{\frac{E}{|\Delta|}
\frac{2L}{\xi_0\cos\theta}}.\\
\end{equation}
The resulting density of bound states for a normal metal length
$L=1.5\xi_0$ is shown in \fref{surfDOS_fig}(b). Note that the density
of surface states in this figure is only the part of the density of
states with $|E|<\Delta_{max}$, similar to the anisotropic $s$-wave
case; there is also a considerable contribution from the extended
states of the bulk due to the gap nodes.

\subsection{Impurity induced MGS}

The above discussion was focused on MGS at planar surfaces. However,
it is clear that MGS must exist at any defect which produces
quasiparticle scattering between states $\theta$ and $\bar\theta$
corresponding to different signs of the order parameter: cracks, twin
boundaries, impurities, etc. Thus, a direct measurement of the DOS at
zero energy by scanning tunneling microscopy (STM) in the vicinity of
an impurity can test the d-wave symmetry of the order parameter. Such
experiments were recently done by \citeasnoun{Yazdani},
\citeasnoun{Hudson} and \citeasnoun{Pan}. We shall discuss these
experiments later, at the end of Sec IV, which is devoted to a
discussion of current transport through MGS.

\subsection{Paramagnetic effect of MGS}

\begin{figure}
\centerline{\includegraphics[width=10cm]{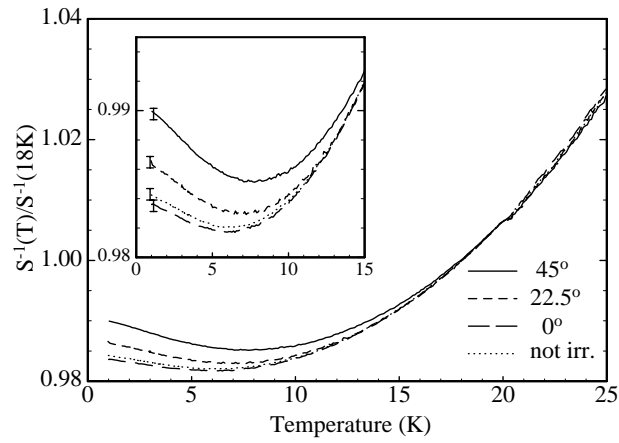}}
\caption{Temperature dependence of the penetration depth $\lambda
  (\propto S^{-1})$ measured on YBCO films containing differently
  oriented linear defects introduced via ion bombardment. Due to the
  presence of MGS at the defects, which respond paramagnetically to an
  external field, there is an upturn of $\lambda$ at low temperature.
  The upturn in the unirradiated film is thought to be due to MGS at
  internal interfaces (grain boundaries). What is important is the
  enhancement of the paramagnetic effect with increasing orientation
  angle, compare the angle $\alpha$ in Fig.~\ref{surface_d_fig}, from
  $0$ to $45^o$ of the defects. From \citeasnoun{Walter}, copyright
  (1998) by the American Physical Society.}\label{Walter_fig}
\end{figure}

Another possibility to detect the impurity induced MGS was
demonstrated in the experiment by \citeasnoun{Walter}
(\fref{Walter_fig}), where the response to an external magnetic field
was studied. In this experiment, the MGS were introduced in a
$c$-oriented YBCO film by creating multiple quasi-1D in-plane defects
(ion tracks) by ion bombardment. By varying the orientation of the ion
beam, and therefore the ion tracks, with respect to the order
parameter, it was possible to control the density of MGS. In the
experiment, the temperature dependence of the magnetic penetration
depth $\lambda(T)$ was measured for different orientations of the
defects (\fref{Walter_fig}). It was found that for (110) orientation
$\lambda(T)$ has a minimum below 15K and increased with decreasing
temperature, while for (100) orientation the low-temperature upturn
was much smaller.

This effect can be explained within the above picture by studying the
paramagnetic response of MGS to an applied magnetic field
\cite{Higashitani}. As was mentioned in \sref{surf_section}, the
surface states undergo splitting in the presence of a magnetic field.
The energy of an individual MGS becomes $E_{MGS}(\theta)=p_F v_{sy}
\sin\theta$, where $v_{sy}$ is the superfluid velocity along the
surface of the defect.  Since the MGS are localized over a distance
much smaller than the penetration depth ($\xi_0\ll\lambda$), the
nonhomogeneity of the superflow can be neglected. The MGS current per
$ab$-plane along the surface is $j_{MGS}=ev_F\sin\theta
|\psi_{MGS}(x)|^2 n_F[E_{MGS}(\theta)]$. The total current of all MGS
at the defect is calculated by integrating over $x$ and also over
$k_y$, $-k_F<k_y<k_F$.  At zero temperature, $T=0$, only the negative
energy levels ($k_y<0$), which carry paramagnetic current, are
occupied, which leads to a total current $j_{MGS}=-eE_F/2\pi\hbar$ at
the defect [for (110) orientation]. By dividing this equation by $d$,
the average distance between the defects, we get the average density
of the paramagnetic MGS current. At finite temperature, the positive
energy levels carrying diamagnetic current also become occupied, which
leads to a reduction of the paramagnetic effect.  When $T\gg p_F v_s
$, expansion of the Fermi function and integration over $k_y$ yield a
$1/T$-dependence for the average MGS current density, $j_{MGS}=
-(eE_F/6\pi\hbar d)( p_F v_s/T)$.  This paramagnetic current is to be
added to the paramagnetic current of the quasiparticles in the bulk of
d-wave superconductor, $j_n= -e v_s n_n(T)$, where $n_n(T)$ is the
density of quasiparticles which linearly depends on the temperature,
$n_n(T)\sim(k_BT/\Delta)n$, at $k_BT\ll\Delta$ ($n$ is the electron
density).

This interplay of the bulk quasiparticle current and the MGS current
produces the low-temperature anomaly of the magnetic penetration depth
observed in the experiment \cite{Walter}. The effect is most
pronounced for (110) orientation, and disappears for (100) orientation
of the defects because of the absence of MGS in the latter case. In
the experiment, the effect did actually not disappear for the (100)
orientation, but was a minimum. The finite value was explained to be
due to MGS at internal interfaces in the film, e.g. grain boundaries.

\subsection{Splitting of MGS}\label{MGSsplit}

Although the $d$-wave symmetry of the gap necessarily leads to the
formation of MGS, the resulting large density of states at the Fermi
level is actually energetically unfavorable: any mechanism able to
split the MGS and produce a gap in the spectrum will lower the energy
of the system \cite{Sigrist,FY}. Different mechanisms splitting the
MGS have been proposed in the literature: (1) splitting due to a
subdominant component of the order parameter which forms a local
complex $d\pm is$ gap parameter at the surface, (2) self-induced
Doppler shifts, and (3) formation of spin-density wave near the
surface.  We shall here discuss the first two mechanisms in greater
detail, while for the third scenario, we refer to the original
literature \cite{HonSig,HonWakSig}.

\subsubsection{$d\pm is$ order parameter}

For orientations where the $d$-wave gap is suppressed and at low
temperatures, a subdominant gap of $s$ or $d_{xy}$ symmetry may appear
near the surface if a weak secondary interaction (additional to the
one responsible for the $d$-wave superconductivity) is present
\cite{MS2,MS3,KTKK,MS4,FRS,Tanuma_PRB99,Zhu_PRB99}. In a
self-consistent picture, the complex combination
$\Delta_{d_{x^2-y^2}}\pm i\Delta_s$ (or $\Delta_{d_{x^2-y^2}}\pm
i\Delta_{d_{xy}}$) is energetically favorable. The MGS is sensitive
to the presence of this subdominant gap; to demonstrate this, let us
consider the order parameter $\Delta_{d_{x^2-y^2}} + i\Delta_s$ with
the $d$-wave component having $\alpha=\pi/4$ orientation. We now
define a relative phase angle $\chi_{rel}=\arctan
(\Delta_s/\Delta_d)$. In this case, the scattering phase shift during
Andreev reflection has the additional contribution $+\chi_{rel}$ for
$k_y>0$, and $-\chi_{rel}$ for $k_y<0$. In a symmetric case with
$\gamma=\bar\gamma$, we get the following modification of
equation~\eref{BS_dd} at low energies:
\begin{equation}\label{t-rev_ss}
-2\gamma \mp2\chi_{rel}+\pi=2n\pi.
\end{equation} 
The effect of this $\pm k_y$-dependent contribution from the relative
phase is to shift the MGS away from the Fermi level
to the edges of the subdominant gap,
\begin{equation}\label{t-rev_MGS}
E_{MGS}=- (\mbox{sgn}k_y)\; \Delta_s,
\end{equation}
and the $\pm k_y$ degeneracy of the MGS is lifted. This splitting of
the MGS is a result of the broken time-reversal symmetry in the
$\Delta_d+i\Delta_s$ state. The MGS splitting drives the instability
(lowers the energy of the system) which appears at a critical
temperature determined by the strength of the subdominant interaction.
Each bound state carries current parallel to the surface in the
direction given by $k_y$, which means that at low temperatures, when
only the state below the Fermi energy is occupied [the one labeled by
$+k_y$ for $\Delta_s>0$ in equation~\eref{t-rev_MGS}], a net surface current
will be present. This spontaneous current manifests the broken
time-reversal symmetry of the order parameter \cite{Sigrist}.  The
TRSB state is two-fold degenerate: the gap parameter can be either
$d+is$ or $d-is$, corresponding to the two possible directions of the
spontaneous surface current.

The spectrum of surface states in $d+is$ superconductors has a close
similarity to the spectrum in a B-phase $p$-wave superconductor. In
the light of the above discussion, it is therefore straightforward to
understand the results obtained by \citeasnoun{pwave_MGS}. The B-phase
gap parameter in a two-dimensional $p$-wave superconductor can be
written $\Delta_{\sigma\sigma} = \Delta(\sigma\cos\theta+i\sin\theta)$
($\sigma=\pm 1$ for down/up spins), see e.g. the book by
\citeasnoun{VollhardtWolfle}. This gap function is similar to the
$d+is$ case above, but with a propagation-angle-dependent magnitude of
the imaginary component $i\Delta\sin\theta$. Thus, there are surface
bound states also for $p$-wave superconductors; however, the bound
state is located at the Fermi level only for quasiparticles
propagating perpendicular to the surface ($\theta=0$), since the gap
is real (and changes sign at the normal reflection event) only for
this particular propagation angle.  Integrating over the injection
angle, we find that the density of states is finite for all subgap
energies, and there is no zero energy peak \cite{pwave_MGS}. We
conclude that the density of bound surface states of $p$-wave and
$d$-wave superconductors are quite different, although the mechanism
of the bound state formation is the same.

\subsubsection{Doppler shifts}

In connection with the discussion of the paramagnetic effect,
\citeasnoun{Higashitani} discovered that the MGS are unstable, even in
the absence of any subdominant interaction (pure $d$-wave order
parameter), due to a possibility of creation of spontaneous magnetic
fields. As we discussed in the previous section, a splitting of MGS
yields a surface current at low temperature. This current creates a
magnetic field which is screened by a supercurrent counterflow which,
in turn, further splits the MGS due to the Doppler shift effect
discussed in \sref{surf_section}. The effect saturates when the
magnetic energy compensates for the energy of the Doppler shifted MGS.
Thus, a time-reversal symmetry breaking (TRSB) state spontaneously
occurs, with a surface current due to the split MGS and a
self-consistent screening current upholding the Doppler shifts at a
phase transition temperature
$k_BT_{TRSB}/\Delta_0=(1/6)(\xi_0/\lambda)$
\cite{HonSig,BKK_prb00,LSW} which is much smaller than the critical
temperature of the type-II superconductors ($\xi_0\ll\lambda$).

\section{dc Josephson effect}

In this Section we shall discuss how, from a theoretical point of
view, MGS influence the dc Josephson effect. First, we develop the
concept of Andreev bound states and explain how they in many cases
determine the properties of Josephson junctions. Then we discuss some
particular junction orientations which illustrate some effects
peculiar for Josephson junctions of $d$-wave superconductors: large
Josephson currents proportional to $\sqrt{D}$ ($D$ is the junction
transparency), $1/T$ dependence ($T$ is the temperature) of the
Josephson critical current, $\pi$-junction formation,
$0\rightarrow\pi$ junction crossovers with temperature, and
spontaneous time-reversal symmetry breaking (TRSB).

\subsection{Andreev states: coupled surface states}

\begin{figure}
\centerline{\includegraphics[width=10cm]{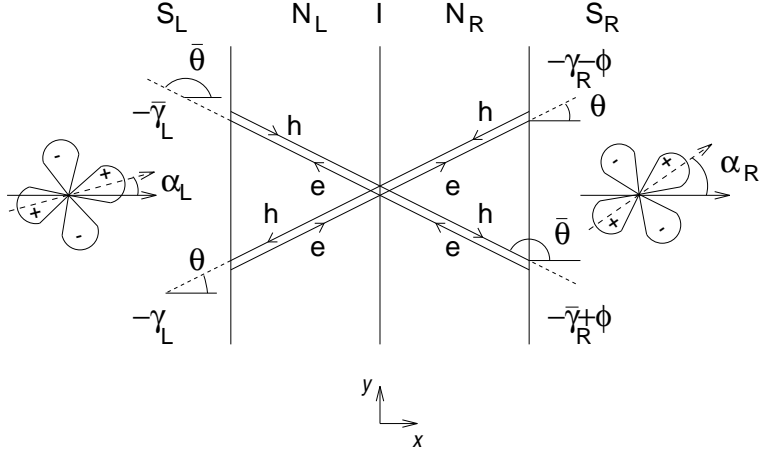}}
\caption{Quasiclassical trajectory illustrating the Andreev bound state
  in a specular two-dimensional superconductor junction. $S_L$ and
  $S_R$ are the left and right $d$-wave superconductor electrodes with
  orientation angles $\alpha_L$ and $\alpha_R$ respectively. The
  junction region consists of a barrier (insulator I) and normal
  regions $N_L$ and $N_R$ which may be due to the gap suppressions.
  The Andreev state can be thought of as a hybridization of the
  surface states (Fig.~\ref{surface_d_fig}) of the left and right
  sides of the insulator.}\label{Andreev_state_fig}
\end{figure}

When two superconducting electrodes are connected to form a junction,
the superconducting surface states form coherent bound states of the
entire junction, Andreev bound states. There is an obvious analogy
with a double well system. The structure of the Andreev state is
illustrated in \fref{Andreev_state_fig}. Like the surface states, the
Andreev state consists of a combination of electron and hole wave
functions with opposite electron momenta $k_x$ (momentum $k_y$
parallel to the specular surface is conserved).  The analogy with the
double well system allows us to expect that the Andreev levels will be
split and shifted with respect to the surface levels, the shift
depending (i) on the transparency of the tunnel barrier separating the
electrodes and (ii) on the superconducting phase difference across the
junction. Taking into account the phase dispersion of the energy
($dE/d\phi$) of the Andreev levels, one may talk about Andreev bands
$E(\phi)$ with width (dispersion) proportional to the junction
transparency $D$.  However, if the surface states at the two sides of
the junction have equal energies, the coupling becomes resonant. In
this {\em resonant} case, the splitting of the levels, and therefore
the width of the Andreev band, will be particularly large and
proportional to $\sqrt D$ \cite{Giant}.  As we will see, the width of
the Andreev band determines the critical Josephson current, and the
Andreev band dispersion determines the Josephson current-phase
relation.  One might expect the surface states to be degenerate in
junctions fabricated with the same superconducting material; however,
even in this case the degeneracy can be lifted due to several types of
asymmetries, e.g.  misorientation of the order parameters, difference
in spatial modulation of the the order parameters, etc.

The derivation of the spectral equation for the Andreev states is more
involved than the derivation of the spectrum of surface states. The
properties of the Andreev levels can therefore not be strictly deduced
from simple quasiclassical arguments but rather require solving the
Bogoliubov-de Gennes (BdG) equation \cite{deG_book}. However, the
spectral equation can be presented in terms of the earlier introduced
phase shifts $\gamma_L$ ($\gamma_R$) and $\beta_L$ ($\beta_R$) in the
left (L) and right (R) electrodes of the junction, and the
superconducting phase difference $\phi=\chi_L-\chi_R$. We here choose
to define the Andreev reflection phase $\gamma$ for negative $\Delta$
as $\gamma=\arccos(E/|\Delta|)+\pi$. The spectral equation then
becomes
\begin{eqnarray}\label{dd_ABS}
\cos a = R(\theta)\cos b + D(\theta)\cos c,
\end{eqnarray}
where
$2a=-\gamma_L-\bar{\gamma_L}-\gamma_R-\bar{\gamma_R}+\beta_L+\beta_R$;
$2b=-\gamma_L-\bar{\gamma_L}+\gamma_R+\bar\gamma_R+\beta_L-\beta_R$;
$2c=-\gamma_L+\bar{\gamma_L}-\gamma_R+\bar{\gamma_R} - 2\phi$.

\Eref{dd_ABS} describes hybridization of the surface states.  This is
obvious if one lets the transparency $D(\theta) \rightarrow 0$, in
which case one recovers equation~\eref{BS_dd} for decoupled surface states.
However, equation~\eref{dd_ABS} can also be interpreted as hybridization of
Andreev states present in the corresponding perfectly transparent
junction ($D=1$) due to normal electron back scattering at the
insulating barrier. The spectral equation for Andreev states in
perfectly transparent junctions actually follows from quasiclassical
arguments:
\begin{eqnarray}\label{dd_ABS_D1}
-\gamma_L-\gamma_R + \beta_L+\beta_R-\phi = 2n\pi,\nonumber\\
-\bar{\gamma_L}-\bar{\gamma_R}+\bar\beta_L+\bar\beta_R-\phi =2n\pi,
\end{eqnarray}
where the first and second lines correspond to the decoupled $\theta$
and $\bar\theta$ segments in \fref{Andreev_state_fig}.

\subsection{Josephson current}

Since Andreev reflection is accompanied by charge transfer through the
NIS interface, the Andreev states are able to carry current. The
current through the surface states is blocked in the direction
perpendicular to the surface due to complete quasiparticle reflection
from the insulator. In the NIS junction, this insulating interface
becomes transmissive, and the current may flow through the Andreev
states.  There is a general relation between the current through the
Andreev state and the phase dispersion of the energy of the Andreev
state,
\begin{equation}
I_x={2e\over \hbar}{dE\over d\phi}.
\end{equation}
This equation can be derived directly from the BdG equation \cite{Sh}
or deduced from the thermodynamical equation $I=(2e/\hbar)\partial
F/\partial \phi$ by using a microscopic expression for the junction
free energy \cite{Beenakker}. The total Josephson current of the bound
states per unit surface area, per $ab$-plane, flowing perpendicular to
the surface is
\begin{eqnarray}\label{bs_current}
j_x &=&\frac{2ek_F}{h}\sum_n \left< \frac{dE_n}{d\phi}n_F(E_n)\right>
\end{eqnarray}
where the angle brackets denote integration over the trajectory angle
$\theta$,
\begin{eqnarray}
\left<...\right>=\int_{-\pi/2}^{\pi/2}d\theta\cos\theta.
\end{eqnarray}

The Josephson current in d-wave junctions strongly depends on the
orientation of the order parameters in the electrodes with respect to
the junction interface. The whole picture is quite complex, however
one can distinguish three qualitatively different cases: (i) no MGS at
any side of the junction, (ii) MGS at both sides of the junction, and
(iii) MGS at one side of the junction. For arbitrary orientation all
three cases may occur for different sectors of the trajectory angle
$\theta$, and the total current is a weighted sum of currents of the
three types. In a pure form, the first case is realized in $d_0/d_0$
junctions where the situation is similar to $s$-wave junctions: no
suppression of the order parameter, and the two Andreev bound states
which dominate the Josephson current are situated close to the gap
edges and have dispersion of order $D$. The second case is realized in
a pure form in $d_{\pi/4}/d_{\pm\pi/4}$ junctions. In this case, the
central phenomenon is the {\em resonant coupling of MGS} which
produces large current proportional to $\sqrt{D}$ with rather special
properties. We choose to illustrate the third case by the
$d_0/d_{\pi/4}$ junction. The peculiarity of this case is the
time-reversal symmetry breaking, TRSB.

In many works, e.g. \cite{SR}, it has been assumed that tunneling
occurs only for normal incidence, $\theta=0$. This is motivated by the
discriminating effect of the {\em tunneling cone} \cite{Wolf}.
However, in order to include the effects of MGS it is very important
actually to take a wide tunneling cone $D(\theta)$ into account, since
tunneling through MGS only occurs at finite angles of incidence.
Ignoring this point, one might conclude that the Josephson current is
zero, or at least small, for the $d_{\pi/4}/d_{\pi/4}$ orientation
since $\Delta(\theta=0)=0$. However, this will be true only for a cone
very strongly peaked around $\theta=0$, since the MGS contribution for
finite cone angle is resonant. Clearly, the exact functional
dependence of the current on orientation will be sensitive to the
actual shape of the cone and not only to the value of the junction
transparency $\left<D\right>$. Throughout this review we use a
$\delta$-function model for the barrier. In this case the cone is
quite wide, $D(\theta)\sim\cos^2\theta$ [see e.g.  \cite{Bruder}], and
the effects of MGS are somewhat overemphasized. An exception is the
treatment of the $d_0/d_{\pi/4}$ orientation below, where we instead
follow reference \cite{LSW} and use a wide barrier model which has
$D(\theta)\sim exp(-\kappa\theta^2)$, where $\kappa$ determines the
width of the cone and is a function of the barrier height and width
\cite{Wolf}.  However, even in this case the MGS contribution will
dominate at low temperature, as discussed below.

\subsubsection{$d_{\pi/4}/d_{\pi/4}$ orientation: $0$-junction with large $I_c$}

The order parameter for this orientation is strongly suppressed by the
surface; however, the qualitative properties of MGS in junctions with
low transparency $D\ll 1$ are not sensitive to the detailed profile of
the order parameter, which allows us to consider the rectangular well
model.

\begin{figure}
\centerline{\includegraphics[width=8cm]{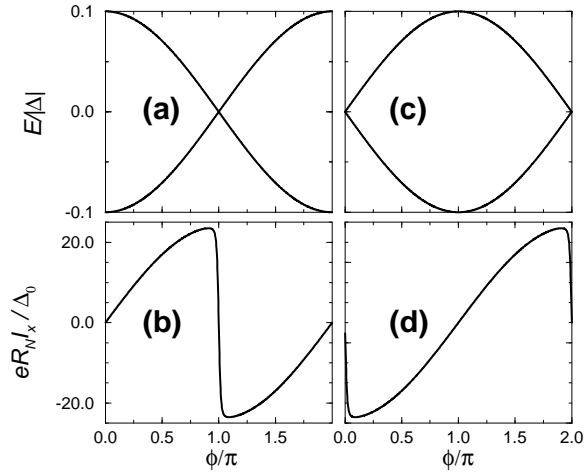}}
\caption{(a) Dispersion relation $E(\phi)=\Delta(\theta)\sqrt{D(\theta)}\cos\phi$
  for the Andreev bound states in the specular $d$-wave junction with
  orientation $d_{\pi/4}/d_{\pi/4}$ for the injection angle
  $\theta=\pi/4$. The transparency is $D(\theta=\pi/4)=0.01$.  (b)
  Angle integrated current-phase relation for the same junction for
  very low temperature ($T=0.003T_c$). (c)-(d) The same but for the
  orientation $d_{\pi/4}/d_{-\pi/4}$. In this case the dispersion
  relation is $E(\phi)=\Delta(\theta)\sqrt{D(\theta)}\sin\phi$ and the
  current is $\pi$-shifted relative to the $d_{\pi/4}/d_{\pi/4}$ case.
  The angle averaged transparency is $<D>=0.026$.}\label{d45d45_fig}
\end{figure}

For this high-symmetry orientation, the four gaps in
equation~\eref{dd_ABS} have equal magnitudes,
$|\Delta_L|=|\bar\Delta_L|=|\Delta_R|=|\bar\Delta_R|$. The $\pi$ shift
at both sides is picked up at $\bar\theta$ (Andreev reflecting against
a negative lobe of the order parameter). The spectral equation takes
the form
\begin{equation}\label{d45d45_ABS}
\cos\left(2\gamma - \beta \right) = -R(\theta) +
D(\theta)\cos\left(\phi\right),
\end{equation}
which, for small energies $E\ll\Delta$ and negligible dephasing
($\beta=0$), has the solution \cite{Bagwell_PRB98}
\begin{equation}\label{bs_d45d45}
E_\pm=\pm|\Delta|\sqrt{D}\cos\frac{\phi}{2},
\end{equation}
plotted in \fref{d45d45_fig}(a). The current of these two MGS is
calculated via equation~\eref{bs_current}. At zero temperature, only the level
below zero energy is populated while the level above is empty, and the
current is
\begin{eqnarray}\label{dc_d45d45}
j_{MGS} = \frac{ek_F}{h}
\left< |\Delta|\sqrt{D} \right>
\sin{\phi\over 2} \;
\mbox{sgn} \left(\cos{\phi\over 2}\right).
\end{eqnarray}
The critical (maximum) Josephson current is proportional to $\sqrt{D}$
\cite{Tanaka_PRB96,Bagwell_PRB98} and is much larger than the
Ambegaokar-Baratoff critical current in non-resonant s-wave tunnel
junctions, which is proportional to $D$ \cite{AB}. In
\fref{d45d45_fig}(b), we plot the current-phase relation at low
temperature. Note that the $\sqrt{D}$-behavior is a general
phenomenon connected with resonance coupling \cite{Giant}.

To understand the basic temperature dependence of the current, we
notice that the two MGS energies $E_\pm$ carry currents in opposite
directions, $j_+=-j_-$, and therefore with increasing population of
the upper level at large temperature, the current carried by MGS will
decrease,
\begin{eqnarray}
j_{MGS}&=&\frac{2ek_F}{h}
\left\langle\frac{dE_+}{d\phi}n_F(E_+)
+\frac{dE_-}{d\phi}n_F(E_-)\right\rangle\nonumber\\
&=&-\frac{2ek_F}{h}
\left\langle\frac{dE_+}{d\phi}\tanh\frac{E_+}{2k_BT}\right\rangle.
\end{eqnarray}
The scale of the temperature dependence of the current is set by the
well-defined energy difference of the two-level MGS system.  When
$E_+\ll k_BT$, the expansion $\tanh x\approx x$ leads to a $1/T$
dependence of the current
\cite{Tanaka_PRB96,Barash_PRL96,Bagwell_PRB98},
\begin{equation}
j_{MGS}=\frac{2ek_F}{h}\frac{\sin\phi}{8k_BT}\langle
D|\Delta|^2\rangle.
\end{equation}

\begin{figure}
\centerline{\includegraphics[width=8cm]{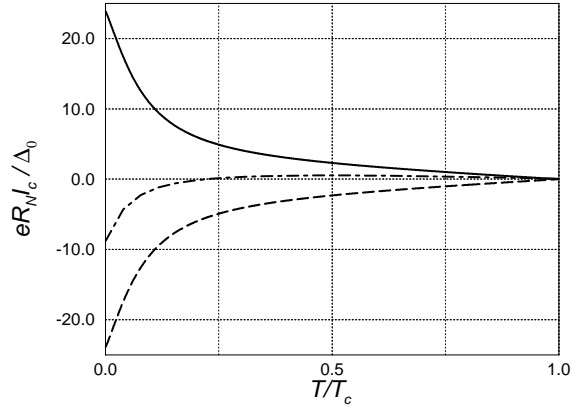}}
\caption{Critical current as a function of temperature for three
  different orientations of the $d$-wave superconductors:
  $d_{\pi/4}/d_{\pi/4}$ (solid line), $d_{\pi/4}/d_{-\pi/4}$ (dashed
  line), and $d_{\pi/8}/d_{-\pi/8}$ (dot-dashed line). The negative
  sign of $I_c$ indicates that the junction minimum is at $\phi=\pi$
  ($\pi$ junction). The $0\rightarrow\pi$ junction crossover with
  decreasing temperature for the $d_{\pi/8}/d_{-\pi/8}$ junction is
  due to the competition between the MGS currents (dominating at low
  temperature) and current contributions from gap-edge bound states
  and continuum states (dominating at high temperature). The angle
  averaged junction transparency is $<D>=0.026$.}\label{Ic_fig}
\end{figure}

In tunnel junctions, the current carried by MGS dominates at low
temperature $k_BT\leq E_+$ because the continuum state current is
proportional to the junction transparency $D$. Due to suppression of
the gap at the interface, there are also Andreev states near the gap
edges which may carry large currents due to degeneracy of the
corresponding surface states. However, the net current of these states
is also of order $D$ since they form pairs far below the Fermi level
which are almost equally populated. The currents carried by the two
bound states of the pair flow in opposite directions and cancel each
other. Thus we conclude that the critical current $I_c$ scales with
temperature as $1/T$ for low temperatures, as clearly seen in
\fref{Ic_fig} (the solid line). In the limit $T\rightarrow 0$, the
$I_c(T)\propto 1/T$ dependence approaches a maximum value
$I_c(0)\propto|\Delta|\sqrt{D}$, c.f. equation~\eref{dc_d45d45}.

\subsubsection{$d_{\pi/4}/d_{-\pi/4}$ orientation: $\pi$-junction with large $I_c$}

From a technical point of view, this case is very similar to the
previous one, the difference being that the $\pi$ shift at the right
side is now picked up at $\theta$ rather than at $\bar\theta$ (because
the positive and negative lobes of the order parameter have changed
places). This results in a phase shift $\phi\rightarrow\phi+\pi$ in
equation~\eref{d45d45_ABS} which now reads
\begin{eqnarray}
\cos\left(2\gamma - \beta \right) =
-R(\theta) - D(\theta)\cos\left(\phi\right),
\end{eqnarray}
with the MGS solution (Riedel and Bagwell, 1998)
\begin{equation}\label{bs_d45dm45}
E_\pm=\pm|\Delta|\sqrt{D}\sin\frac{\phi}{2}.
\end{equation}
The MGS band again has a large dispersion proportional to
$\Delta\sqrt{D}$; however, the phase dependence is $\pi$ shifted
compared to the $d_{\pi/4}/d_{\pi/4}$ junctions [see
\fref{d45d45_fig}(c)]. The current-phase relation will therefore also
be $\pi$-shifted [c.f.  \fref{d45d45_fig}(d)] which puts the minimum
of the junction energy at a phase difference equal to $\pi$ instead of
$0$, corresponding to a $\pi$-junction. This fact was used in the
tri-crystal ring experiments, although the theoretical idea behind the
experiment put forward by \citeasnoun{Larkin} and later on by
\citeasnoun{SR} was based on symmetry arguments and not on the
microscopic picture outlined here.

Again, the critical current has the low-temperature anomaly,
$I_c(T)\propto 1/T$ and approaches a maximum value
$I_c(0)\propto|\Delta|\sqrt{D}$ at zero temperature, (dashed line in
\fref{Ic_fig}). To emphasize the $\pi$-junction behavior, we let the
critical current be negative in this case.

\subsubsection{$d_{\alpha}/d_{-\alpha}$ orientation: $0$ to $\pi$-junction cross-over}

We are now able to discuss general properties of the Josephson current
in symmetric junctions $d_{\alpha}/d_{-\alpha}$; see
\cite{Barash_PRL96,Tanaka_PRB96,Tanaka_PRB97,Tanaka_super} for a
detailed discussion. In these junctions, MGS exist only within certain
limited intervals of the angle $\theta$ while for other angles, the
Andreev level spectrum is similar to the one in s-wave junctions and
the corresponding currents follow the Ambegaokar-Baratoff law. At low
temperature, the MGS will dominate and the equilibrium phase
difference is, as in the $d_{\pi/4}/d_{-\pi/4}$ case, at $\phi=\pi$.
However, at larger temperatures this contribution will be suppressed
and the non-MGS current may dominate. This $s$-wave type current will
give a minimum of the junction energy at $\phi=0$.  Hence, with
increase of the temperature, the junction will undergo a transition
from a $\pi$-junction to an ordinary junction
\cite{Barash_PRL96,Tanaka_PRB96}. The critical current is zero at the
transition temperature where there is detailed balance between the MGS
current and the $s$-wave like current. We illustrate this behavior
with the $d_{\pi/8}/d_{-\pi/8}$ junction, the dot-dashed line in
\fref{Ic_fig}.

\subsubsection{$d_{\pi/4}+is\,/\,d_{-\pi/4}+is$ orientation: $0$ junction re-entrance}

\begin{figure}
\centerline{\includegraphics[width=8cm]{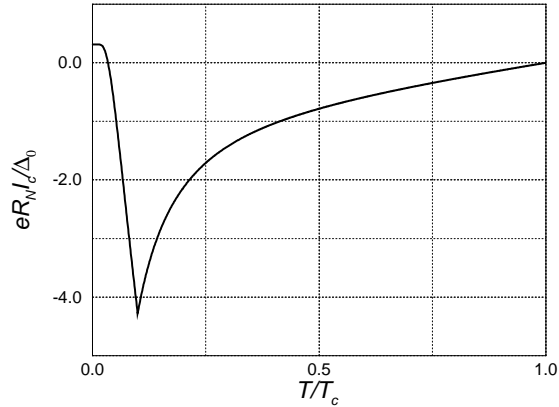}}
\caption{Critical current for the $d_{\pi/4}/d_{\pi/4}$ orientation.
  At low temperature, $T_s$, a subdominant component of the order
  parameter with $s$-wave symmetry is assumed to appear and be $\pi/2$
  out of phase with respect to the dominant $d$-wave component. This
  results in a cut off of the MGS contribution and the junction
  changes from a $\pi$-junction to $0$-junction when the temperature
  is decreased.}\label{Ic_complex_op}
\end{figure}

Development of a subdominant s-wave order parameter near the junction
interface leads to an interesting modification of the low-temperature
anomaly of the critical current. As we earlier discussed, in the
presence of a complex order parameter $\Delta_{d} + i\Delta_s$ the MGS
are split into two states $ E_{MGS}=\pm \Delta_s$, where the sign
depends on the sign of the $k_y$ wave vector. Thus the Andreev
spectrum presented in \fref{d45d45_fig} will move either up or down
(for opposite signs of $k_y$) as soon as the subdominant gap
$\Delta_s$ appears and grows with decreasing temperature. Let us
consider the limit $\Delta_s(0)\gg |\Delta|\sqrt D$ where the effect
is most pronounced.  In this case, the Andreev bands do not cross the
Fermi level at $T=0$; the lower pair of levels is then fully occupied,
while the upper one is empty and therefore does not contribute to the
current.  This leads to an uncompensated $y$-current along the
interface. Large $x$-currents carried by the occupied levels flow in
opposite directions and cancel each other, leading to a small (for
$D\ll 1$) residual current proportional to the first power of the
junction transparency $D$.  This residual current has positive sign,
similar to the common situation in $s$-wave SIS and SNS junctions
\cite{Giant,Samuelsson}, and therefore the $\pi$-junction property is
lost. Thus the Josephson current undergoes a crossover from
$\pi$-junction to $0$-junction when the temperature decreases, and the
critical current has a non-monotonic temperature dependence and turns
to zero at the crossover point \cite{TanKas_sub}, see
\fref{Ic_complex_op}. Experimental observation of such {\em reentrant}
temperature dependence of the critical current would be a direct
evidence of the subdominant order parameter.

\subsubsection{$d_0/d_{\pi/4}$ orientation: $\pi$-periodicity}

For the $d_0/d_{\pi/4}$ orientation, there are MGS only at one side of
the junction corresponding to the non-resonant case. The spatial
shapes of the order parameters are different at the two sides of the
junction which also leads to non-resonant coupling of the non-MGS
surface states. To understand the type of bound states present in this
junction, we solve the spectral equation for the injection angle
$\theta=\pi/8$ when all four $d$-wave gaps involved have equal
magnitudes $|\Delta(\pm\pi/8)|=\Delta_0/\sqrt 2$. The spectral
equation then takes the form [$D=D(\pi/8), \Delta =\Delta(\pi/8)$]
\begin{equation}
\sin(2\gamma-\beta)=\pm D\sin\left(\phi\right)
\end{equation}
where the different signs correspond to quasiparticle states with $\mp
k_y$. The solution for $\beta=0$ is given by
\begin{eqnarray}\label{bs_d0d45}
E^2 &=& \frac{\Delta^2}{2}\left (1\pm\sqrt{1-D^2\sin^2\phi}\right).
\end{eqnarray}
There are two Andreev bands for a given $k_y$, the MGS band and the
band at the gap edge (Riedel and Bagwell, 1998). The dispersion of the
MGS band is proportional to $D$ at low transparency,
\begin{equation}\label{bs_d0d45'}
E_{MGS}  = -\mbox{sgn}k_y\frac{\Delta}{2}D\sin\phi + \mbox{O}(D^2),
\end{equation}
while the dispersion of the Andreev band at the gap edge is much
smaller, proportional to $D^2$. \Eref{bs_d0d45'} is a good
approximation for all injection angles in the low transparency limit
when the MGS are well inside the gap, close to zero.

\begin{figure}
\centerline{\includegraphics[width=8cm]{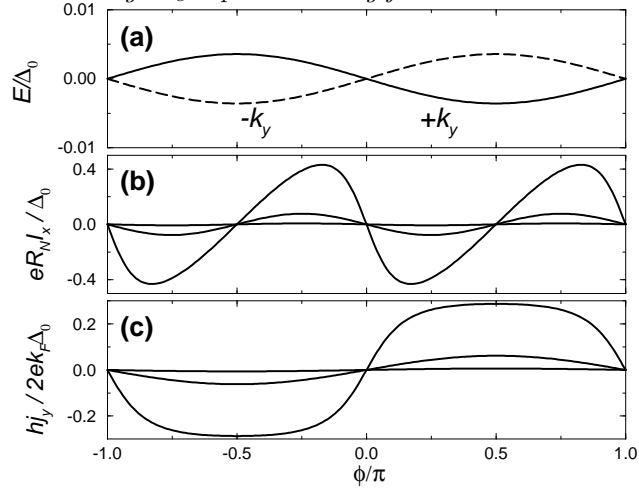}}
\caption{(a) Andreev bound states in a specular $d$-wave junction with
  the $d_0/d_{\pi/4}$ orientation for the injection angle
  $\theta=\pi/9$ and transparency $D(\theta=\pi/9)\approx0.01$. The
  solid and dashed lines are the $+k_y$ and $-k_y$ states
  respectively. (b) The angle integrated current-phase relation for
  this orientation at three different temperatures:
  $T=\{0.001,0.01,0.1\}T_c$. (c) Phase dependence of the surface
  current density calculated to the right of the barrier for the same
  three temperatures. As the temperature is decreased, the $\pm k_y$
  MGS become unequally populated and start to dominate the Josephson
  current which then is enhanced and is of order $D$. At the same time
  the surface current becomes appreciable.}\label{d0d45_fig}
\end{figure}

For the orientations discussed in the previous sections, the MGS
spectrum is degenerate with respect to $\pm k_y$: in other words,
there is a pair of MGS, $E_\pm$, for every $k_y$. In the present case,
however, this degeneracy is lifted as emphasized in
\fref{d0d45_fig}(a). Inspection of the MGS wave functions for
phase differences between $0$ and $\pi$ shows that, at large distance
from the interface, the $E_-$ wave function decays only for $+k_y$,
while for $-k_y$ it grows exponentially. For the $E_+$ wave function
the situation is the opposite: the wave function decays for $-k_y$ and
grows exponentially for $+k_y$.  This difference between the two
time-reversed states labeled by $\pm k_y$ is a consequence of the sign
difference between the $d$-wave lobes and is the key point in
understanding the possibility of spontaneous time-reversal symmetry
breaking (see below).

The current in the $d_0/d_{\pi/4}$ junction is $\pi$-periodic, as
opposed to the usual $2\pi$-periodicity, as was first theoretically
predicted in \cite{Yip1,Yip2,Tanaka_PRB97,Zago,Ost} and later on
measured in experiments by \citeasnoun{Ilichev3}. For low temperature
the MGS band dominates the current-phase relation plotted in
\fref{d0d45_fig}(b) and the bound states in equation~\eref{bs_d0d45}
give $I_c(T)\propto D^2|\Delta|^2/k_BT$.  However, for large
temperatures $I_c$ will be negligibly small already for temperatures
of the order $|\Delta|D$, in contrast to the $d_{\pi/4}/d_{\pi/4}$
case where this happens for temperatures of the order
$|\Delta|\sqrt{D}$.  This difference is a result of the larger energy
scale given by the large dispersion of the bound states
($\propto\sqrt{D}$) in the resonant $d_{\pi/4}/d_{\pi/4}$ case
compared to the much smaller dispersion ($\propto D$) in the
non-resonant $d_0/d_{\pi/4}$ case.

\subsubsection{$d_0/d_{\pi/4}$ orientation: TRSB}

The current through the junction is zero for the phase differences
$\phi=n\pi/2$. However, the minimum junction energy is achieved for
$\phi_{eq}=\pm\pi/2$, since the MGS energy is the lowest for these
phase differences. Due to the uneven occupation of MGS with opposite
signs of $k_y$, the current parallel to the interface is finite even
in equilibrium, see \fref{d0d45_fig}(c). For example at
$\phi_{eq}=\pi/2$, only MGS with positive $k_y$ are occupied and a
surface current is flowing in the positive $y$-direction. The opposite
happens for $\phi_{eq}=-\pi/2$. Remembering that the state with $-k_y$
is obtained from the $+k_y$ state by the time-reversal operation, and
noting that there is no way of taking the time-reversed state to the
original one by adding an integer number of $2\pi$ to $\phi$, we must
conclude that time-reversal symmetry is spontaneously broken at the
$d_0/d_{\pi/4}$ junction. This fact is manifested by the finite
surface current, which produces a magnetic field at the junction.

The Josephson energy gain for $\phi_{eq}=\pm\pi/2$ is partly
compensated by the energy cost of setting up the Meissner screening
currents which ensure that the spontaneous magnetic field vanishes in
the bulk superconductors far from the junction. By calculating the
thermodynamic potential for the junction system \cite{LSW} it can be
shown that the TRSB state really is favorable. If the orientation is
not exactly $d_0/d_{\pi/4}$, the equilibrium phase difference
$\phi_{eq}$ is shifted from $\pm\pi/2$. However, the TRSB effect (i.e.
$\phi_{eq}\neq 0$ or $\pi$) remains for deviations up to about $10^o$
\cite{YipLTP,LSW} from the $d_0/d_{\pi/4}$ orientation.

Since there are screening currents flowing, we may expect Doppler
shift effects. For transparencies $D\gg\xi_0/\lambda$, the dispersion
of the MGS with phase difference, given by equation~\eref{bs_d0d45'}, is much
larger than the Doppler shifts; these can therefore be neglected and
the picture outlined above holds. However, for $D\ll\xi_0/\lambda$ the
Doppler shifts dominate, and the problem is formally the same as MGS
at a free surface discussed in \sref{MGSsplit}.

The above discussion of the $d_0/d_{\pi/4}$ junction, including the
crossover to the doppler shift driven TRSB surface instability,
closely follows the one presented in the paper by \citeasnoun{LSW}.
The possibility of TRSB in purely $d$-wave junctions was first pointed
out by \citeasnoun{Yip_PRB95} for the pin hole geometry, which is
equivalent to having $D=1$. Many people, using different theoretical
tools, have considered the $d_0/d_{\pi/4}$ junction: by using
Ginzburg-Landau theories \cite{SBL,Ost} or quasiclassical Green's
function techniques \cite{Yip_PRB95,YipLTP,FYK,FY}, or by solving the
Bogoliubov-de Gennes equation \cite{Huck,Zago,ZagoOsh,LSW}.  However,
the final results are qualitatively the same: $\pi$-periodic
current-phase relation and the possibility of spontaneous
time-reversal symmetry breaking. There is however one important
difference: the absence (as above) or presence of a subdominant
component of the order parameter at the junction. We conclude that
broken time-reversal symmetry at a $d_0/d_{\pi/4}$ Josephson junction
will not give any evidence for the presence of a subdominant
component, since the $d$-wave order parameter alone enforces broken
symmetry.

In experiments reported by \citeasnoun{Kirtley} and
\citeasnoun{Mannhart_PRL96} fractional fluxes were found at
grain-boundaries with the $d_0/d_{\pi/4}$ orientation. If
time-reversal symmetry is broken as described above, spontaneous
surface currents offer a possible explanation for these experiments,
see \cite{Sigrist}. However, \citename{Mannhart_Zphys96}
\citeyear{Mannhart_PRL96,Mannhart_Zphys96} proposed an explanation in
terms of randomly distributed facets along the grain-boundary, where
each facet may have equilibrium phase difference $0$ or $\pi$
depending on whether zero, one or two lobes with negative sign of the
$d$-wave order parameter point towards the junction. When the phase
difference varies along the grain-boundary, formation of fractional
fluxes may be energetically favorable. It should be noted that this
model assumes a simple $\sin\phi$ dependence of the current-phase
relation, and does not take into account the formation of MGS at the
junction.

\subsubsection{$d_0/d_0$ junctions with resonant levels in the barrier}\label{Ic_res}

There is an interesting similarity between the Josephson effect in the
$d_{\pi/4}/d_{\pi/4}$ junctions and in $d_0/d_0$ junctions with
resonant states in the tunnel barrier. In the presence in the tunnel
transparency of a narrow Breit-Wigner resonance close to the Fermi
level, the Andreev level spectrum has the form found in the $s$-wave
case \cite{Beenakker_SET,WendinSXS},
\begin{equation}\label{bs_res}
E= \pm\sqrt{E_0^2+\Gamma^2\cos^2(\phi/2)},
\end{equation}
where $E_0$ is the position and $\Gamma$ is the width of the
resonance, $E_0, \Gamma\ll\Delta$. If the resonance is exactly at the
Fermi level, $E_0=0$, the spectrum becomes
\begin{equation}\label{ss_res}
E= \pm\Gamma\cos(\phi/2),
\end{equation}
which is the spectrum of an SNS-junction ($D=1$) with
$|\Delta|=\Gamma$, having dispersion $\Gamma$ instead of $|\Delta|$.
At zero temperature, the current is then given by
\begin{equation}\label{ss_rescurr}
j = \frac{e}{\hbar}\Gamma\sin\frac{\phi}{2} \;
\mbox{sgn}\left(\cos\frac{\phi}{2}\right).
\end{equation}

The temperature dependence of the critical current in this case is
then also $I_c(T)\propto 1/T$ for $k_BT\gg\Gamma$. Thus, while
interpreting experimental results it is useful to have in mind that
the low-temperature anomaly of the critical current is a fingerprint
of a resonant state at the Fermi level rather than indication of the
$d$-wave symmetry, although this resonance (MGS) could be due to the
$d$-wave symmetry of the order parameter.

\subsection{Critical Josephson current in grain boundary and ramp-edge junctions}\label{Ic_exp}

An important class of experimental studies involves the determination
of the critical Josephson current in current-biased artificial grain
boundary (GB) junctions $d_{\alpha_L}/d_{\alpha_R}$ by varying the
orientation angles $\alpha_L$ and $\alpha_R$ of the order parameters
\cite{GrossReview,DelinKleinsasser,Ivanov_prb98,Arie2000}.

Consider a {\it gedanken} experiment where we may freely rotate the
order parameters of the left ($\alpha_L$) and right ($\alpha_R$)
superconducting electrodes, and also independently vary the
transparency $D$ of the interface barrier (I)
(figure~\ref{Andreev_state_fig}), creating ideal
$d_{\alpha_L}/d_{\alpha_R}$ junctions: how will the Josephson current
develop in different situations?

Starting with an ideal $d_0/d_0$ GB junction, there is no MGS, and the
Josephson current $\propto D$ is due to non-resonant transport through
the interface barrier. If we now rotate the order parameter
symmetrically away from $\alpha=0$ to $\alpha=\pi/4$, keeping $D$
fixed, the $d_{\alpha}/d_{-\alpha}$ the junction will develop MGS, and
the critical current will then cross over from $D$ to $\sqrt{D}$
dependence because of resonant transport through MGS. The resulting
current enhancement can become very large ("giant Josephson current")
if the transparency is low, $D\ll 1$, and the $I_cR_N$ product will be
enhanced by $1/\sqrt{D}$ over the Ambegaokar-Baratoff limit
characterizing $d_0/d_0$ tunnel junctions.

In reality one must rotate the order parameter by rotating the
lattice. This necessarily leads to an interface barrier, the strength
of which increases with misorientation angle (up to $2\alpha=\pi/4$
for tetragonal symmetry). Assuming that the thickness of the barrier
increases with misorientation angle, the result will be exponentially
decreasing transparency and narrowing of the tunneling cone.  With
decreasing transparency, the low-temperature region where MGS are
important will decrease ($k_BT \propto \sqrt{\left<D\right>}\Delta_0$
for resonant MGS). These effects may seriously reduce the weight of
the MGS contribution. Therefore, in real experiments, the role of MGS
will depend on the detailed properties of the junction.

It should be noted that if we continue to rotate the structure with
the order parameters to reach the $d_{\pi/4}/d_{-\pi/4}$ limit, there
is no longer any lattice mismatch, and therefore no interface barrier.
The full presence of MGS therefore cannot be achieved in this type of
experiment: to really observe the effects of MGS, one must fabricate
$d_{\pi/4}/I/d_{-\pi/4}$ junctions with separately controlled tunnel
barriers.

In real high-$T_c$ GB junctions, the $I_cR_N$ product decreases
strongly with increasing misorientation angle. In the absence of MGS,
one would expect this product to be constant, but it is experimentally
often found that $I_cR_N \propto 1/\rho_N$, where $\rho_N$ is the
normal resistance per unit area \cite{GrossReview}. An explanation,
due to \citeasnoun{GrossReview}, is that Coulomb blockade prevents
resonant Josephson current but allows resonant normal current
transport through the levels in the barrier. We suggest that this
behavior can also be explained in terms of a constant spectral
density of normal levels with resonance widths $\Gamma$ inside the
barrier. From the discussion in \sref{Ic_res}, we know that for a
resonance with energy $E_0\ll\Gamma$, equation (\ref{ss_rescurr})
gives a resonantly enhanced Josephson current $I \propto \Gamma$,
while for resonances with $E_0\gg\Gamma$ the current is negligibly
small. The contribution to the total current then comes from a
spectral region of width $\Gamma$ around $E=0$, with the results $I
\propto \Gamma^2$. The normal current, on the other hand, is
resonantly enhanced in the entire spectral region, giving $I_cR_N
\propto\Gamma\propto 1/\rho_N$.

In a recent paper, \citeasnoun{Arie2000} have investigated the
properties of the Josephson current in ramp-edge junctions with
$\alpha_R-\alpha_L= \pi/4$ overall misorientation angle, varying the
orientation of the interface (and therefore $\alpha_R,\alpha_L$) to
vary the role of MGS.  In particular, the dependence of the critical
current on temperature has been experimentally investigated and fitted
to d-wave theoretical results for short $d/d$ junctions with high
transparency, D=0.5-0.8, which implies that the dispersion of the MGS
is a large fraction of the gap. To get agreement with experiment for
$T\rightarrow 0$, \citeasnoun{Arie2000} introduced life-time
broadening of the MGS (describing effects of rough interfaces and
inelastic scattering) via an imaginary part of the energy, which
strongly reduces $I_c(T)$ and the $I_cR_N$ product at low
temperatures, changing the sign of the curvature and leading to
saturation of $I_c(T)$.  The final good qualitative agreement between
theory and experiment is consistent with the presence of life-time
broadened MGS in a high-transparency junction.

It should be noted, however, that even if the transparency for
superconductive transport seems high, an estimate of the experimental
{\em average} transparency, based on the junction resistance
($0.2-0.8\Omega$) and area (number of planes and transport channels),
gives a fairly low value, around $D=10^{-3}-10^{-4}$.  We suggest that
the simplest explanation for this difference is that transport takes
place through highly transparent point contacts in the junction
barrier.  In this picture, the temperature dependence of the critical
Josephson current becomes a measure of the true transparency
($D\approx1$), which together with the normal resistance also becomes
a measure of the effective transport area of the junction.

\begin{figure}
\centerline{\includegraphics[width=8cm]{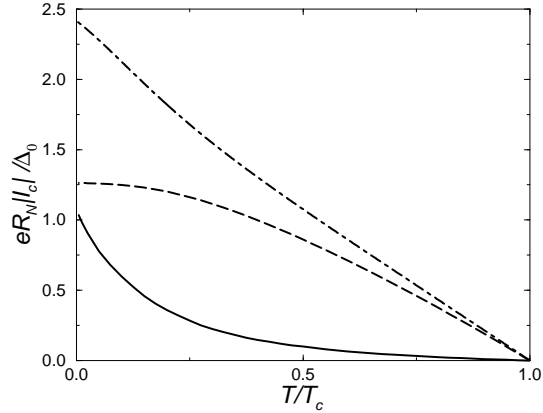}}
\caption{$I_c(T)$ dependences for high transparency, $\left<D\right>=0.68$
  ($Z=0.5$), for three orientations: $d_0/d_0$ dashed line,
  $d_{\pi/4}/d_{-\pi/4}$ dash-dotted line, and $d_{-15^o}/d_{30^o}$
  solid line. The last orientation corresponds to the nominal
  orientation in the experiment by
  \citeasnoun{Arie2000}.}\label{high_D}
\end{figure}

For a more definitive analysis it is desirable to compare experimental
and theoretical results for a larger set of misorientation angles and
transparencies.  To this end we have calculated $I_c(T)$ for high
transparency, $\left<D\right>=0.68$ [$Z=0.5$, as in the original fit
by \citeasnoun{Arie2000}], neglecting MGS broadening. We have
considered the high-symmetry orientations $d_0/d_0$ and
$d_{\pi/4}/d_{-\pi/4}$, as well for the intermediate orientation
$d_{-15^0}/d_{30^0}$ considered by \citeasnoun{Arie2000}. The result
is shown in \fref{high_D}. In the case of $d_0/d_0$ orientation (no
MGS) we find that $I_c(T)$ has a weak negative curvature, while in the
case of $d_{\pi/4}/d_{-\pi/4}$ orientation (MGS is present) we find a
weak positive curvature of $I_c(T)$. Thus, within our model the
positive curvature is characteristic for the presence of MGS, while
the negative curvature is characteristic for the absence of MGS. The
strong positive curvature in the case of the $d_{-15^0}/d_{30^0}$
orientation in \fref{high_D} may look a bit surprising: in fact, only
at low temperatures does the $1/T$-dependence signify MGS; at higher
temperatures it results from a competition between the nearly-linear
MGS contribution and contributions of opposite sign from the continuum
and gap-edge states. Thus, theoretically, positive curvature of
$I_c(T)$ at low temperatures is a fingerprint of MGS.  However, such a
picture needs to be verified by systematic experimental and
theoretical investigations, and the picture may be complicated. For
example, the recent studies of GB junctions by \citeasnoun{Tafuri1999}
show clear positive $I_c(T)$ curvatures for nominally $d_0/d_0$
orientations, and such behavior is also evident in the review by
\citeasnoun{DelinKleinsasser}, perhaps due to orientational disorder
like meandering and faceting.

\citeasnoun{Arie2000} also studied the magnetic field dependence of
the critical Josephson current, finding evidence for $\sin(2\phi)$
behavior.  This might be evidence for $d_0/d_{\pi/4}$ symmetry, but
may also be due to the high transparency (saw-tooth-like dependence of
the $I(\phi)$ current-phase relation).

We conclude that so far, the role of well-defined MGS is difficult to
identify with certainty in studies of the critical Josephson current
in d/d grain-boundary or ramp-edge junctions, although
\citeasnoun{Arie2000} most likely have taken an important step in the
right direction.  In order directly to observe the presence of MGS one
may have to use low-transparency tunnel junctions and utilize
energy-resolving probes, e.g looking for MGS in the form of
characteristic features in I-V characteristics of voltage biased
junctions.

\section{Voltage-biased N/d junctions}\label{Nd_sec}

Since the discovery of tunneling in NS junctions \cite{Giaever}, and
the explanation of the phenomenon on the basis of the tunnel
Hamiltonian model \cite{Bardeen,CFP}, it is commonly accepted that the
tunneling conductance essentially measures the superconductor density
of states (DOS). Hence, tunnel spectroscopy provides a direct means of
detecting the MGS, which should show up as a large conductance peak
centered at zero bias. This zero-bias conductance peak (ZBCP) has
indeed been observed; see
\cite{Geerk,Kash95,Cov,Ekin,Alff_Rapid,WYGS,ACPNG,Alff_prb,Alff_Eur,Ng,Suzuki,Krupke,Aprili,WanWang,CovGreene}
and references therein. See also the recent reviews by
\citeasnoun{Alff_super} and \citename{Kash_super}
\citeyear{Kash_super,Kash_prog}.

We start this section by discussing the theory of tunneling through
MGS. We will discuss two different transport mechanisms: direct
tunneling, and tunneling involving Andreev reflection. The tunnel
model is usually exploited for calculation of direct tunneling, while
Andreev reflection is usually described by means of scattering theory.
We will compare the two approaches and show that depending on the
relation between the junction transparency and the MGS intrinsic
broadening, the MGS resonance may appear either in the single particle
current or in the pair current. In the end we discuss tunnel
experiments on NS junctions and also the recent STM investigations of
impurity bound states.

\subsection{Theoretical model}

In the tunnel model, the angle-resolved local density of states (LDOS)
$\rho_L$ and $\rho_R$ are calculated on the left(L) and right(R) sides
of the junction, and then the tunnel formula
\begin{equation}\label{tunnel_formula}
\fl
I(V,\theta)\propto\int_{-\infty}^{\infty}dE
D(\theta)\left[n_F(E)-n_F(E+eV)\right]
\rho_L(\theta,E)\rho_R(\theta,E+eV),
\end{equation}
is employed to calculate the current. The total current is obtained by
angle integration $I(V)=\left<I(V,\theta)\right>$, and the
conductance is then found by differentiation with respect to the bias
voltage $V$. Since the density of states (DOS) in the normal metal is
approximately energy independent near the Fermi surface, in the zero
temperature limit the conductance will be reduced to
$\left<D\rho_s(eV)\right>$, where $\rho_s$ is the LDOS in the
superconductor.

For the $\alpha=0$ orientation of the order parameter, there is no
MGS and the angle resolved LDOS is the same as in the bulk
$\rho_s\propto|E|/\sqrt{E^2-|\Delta|^2}\Theta(|E|-|\Delta|)$ [see e.g.
references~\cite{MS1,BGZ}]. The tunnel formula gives a conductance
which in the zero temperature limit is simply the bulk DOS corrected
by the angle dependent transparency $D=D(\theta)$ of the tunnel
barrier,
\begin{equation}
G(V)\propto\left<D\frac{eV}{\sqrt{eV^2-|\Delta|^2}}
\Theta(eV-|\Delta|)\right>.
\end{equation}

For other orientations, the conductance will measure LDOS at the
surface rather than the bulk DOS. For the $\alpha=\pi/4$ orientation,
neglecting gap suppression and surface roughness, the LDOS is given by
\cite{MS1}
\begin{equation}
\rho_s(\theta,E)\propto\frac{\sqrt{E^2-\Delta^2}}{|E|}\Theta(|E|-|\Delta|)
+\pi|\Delta|\delta(E).\label{LDOS_MGS}
\end{equation}
The high spectral weight at the gap edges has moved down to zero
energy and formed the MGS, and the tunnel conductance will contain a
peak at zero voltage. The delta-function singularity is usually
removed by introducing broadening via an imaginary part $\eta$ of the
energy representing some relaxation mechanism. This damping $\eta$
determines the width of the MGS resonance.

\begin{figure}
\centerline{\includegraphics[width=8cm]{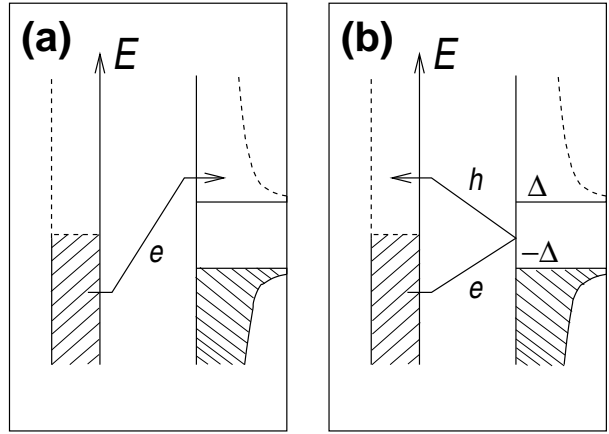}}
\caption{Illustration of (a) the single-particle (quasiparticle)
  and (b) the two-particle (Andreev) processes in the NIS junction. At
  zero temperature allowed processes connect occupied states below
  zero energy (the Fermi level) in the normal metal with (a)
  unoccupied states above the gap in the superconductor, and (b) above
  zero energy in the normal metal. We have aligned the Fermi energies
  of the two sides by a gauge transformation in the right
  superconductor; consequently, quasiparticles are accelerated by $eV$
  at each passing through the junction region.}\label{processes}
\end{figure}

This interpretation of the conductance spectra in terms of
single-particle (quasiparticle) current, see
\fref{processes}(a), assumes that all states in the electrodes
form reservoirs, i.e. their relaxation times are assumed short enough
to ensure equilibrium. This assumption is reasonable for the extended
states of the continuum, while for the MGS, lying deep within the
superconducting gap, it requires special consideration. If the MGS are
true bound states without coupling to the bulk of the superconductor,
i.e. intrinsically very narrow, then current transport through the MGS
is only possible via the Andreev process, see
\fref{processes}(b). The equation for the current then reads
\begin{equation}\label{BTK}
I(V)=I_1(V)+I_2(V),
\end{equation}
where $I_1$ and $I_2$ are the single-particle (quasiparticle) and
two-particle (Andreev) currents respectively. This case of narrow MGS
is most easily studied within the scattering approach
\cite{landauer,buttiker,imry}.

\begin{figure}
\centerline{\includegraphics[width=8cm]{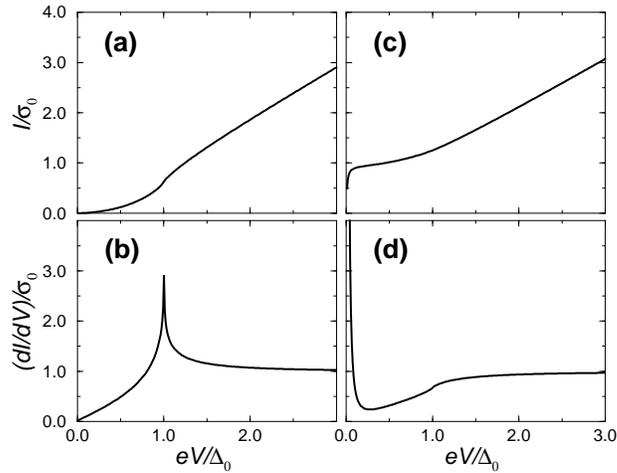}}
\caption{Current voltage relations and corresponding differential
  conductances for the $N/d_{\alpha}$ junction with two different
  orientations: $\alpha=0$, shown in (a)-(b), and $\alpha=\pi/4$,
  shown in (c)-(d). The zero-bias anomaly in (c)-(d) is due to the MGS
  resonance. The barrier is modeled by a $\delta$-function potential
  with a large tunnel cone with angle averaged transparency
  $<D>=0.026$ which emphasize the importance of the MGS resonance
  which appear for finite injection angles $\theta$ only. The
  temperature is zero and $\sigma_0$ is the normal state
  conductance.}\label{Nd_IV}
\end{figure}

\citeasnoun{BTK} were the first to use scattering theory to calculate
the conductance of an $s$-wave NS junction. Later on,
\citeasnoun{Bruder}, \citeasnoun{Kash95} and \citeasnoun{Ting}
generalized the theory to the $d$-wave case; see also
\cite{Tanaka_PRL95,Kash_PRB96,Tanaka_LDOS_PRB96}. At zero temperature,
in the low-transparency limit, we have for the $\alpha=\pi/4$
orientation
\begin{equation}
I_1(\theta,E)=D\frac{\sqrt{E^2-\Delta^2}}{|E|}\Theta(|E|-|\Delta|),
\end{equation}
and the two-particle current takes the form
\begin{equation}
I_2(\theta,E)=\frac{2 D^2}{D^2 + 4R (E/\Delta)^2}.
\end{equation}
The total current is obtained by integration over angles and energies
\begin{equation}
I(V)\propto\left<\int_{-\infty}^{\infty}dE
\left[n_F(E)-n_F(E+eV)\right]\left[I_1(\theta,E)+I_2(\theta,E)\right]\right>
\end{equation}
The two-particle current contains a Breit-Wigner resonance of width
$\Delta D/2\sqrt{R}$, the MGS, and $I_2\propto \Delta D$ as soon as
$eV>\Delta D$. Hence, the two-particle current is resonantly enhanced
near zero bias (it is of the same order as the single-particle
current), which produces a zero-bias conductance peak. The
current-voltage and conductance-voltage relations calculated within
the scattering approach are plotted in \fref{Nd_IV}.

Consider now the ratio $q$ between the life time due to leakage to the
normal reservoir through the barrier, $\tau_b\propto \hbar/\Delta_0
D$, and the life time due to inelastic relaxation, $\tau_r\propto
\hbar/\eta$, which couple the MGS with the bulk superconductor,
\begin{equation}\label{tau_quota}
q = \frac{\tau_b}{\tau_r} \propto \frac{\eta}{\Delta_0 D}.
\end{equation}
Clearly this life-time ratio will determine whether the MGS will
participate in current transport via the single-particle process (for
large damping; $q\gg 1$), or via the two-particle process (for small
damping; $q\ll 1$).

The fingerprint of Andreev transport is the so-called excess current
defined as \cite{Artemenko}
\begin{equation}\label{excess_current}
I_{exc}=\lim_{V\rightarrow\infty}\left[I_S(V)-I_N(V)\right],
\end{equation}
where $I_S$ and $I_N$ are the currents in the superconducting and
normal states respectively. Following the work of \citeasnoun{BTK}, it
became common practice to estimate the efficiency of Andreev
reflection from the magnitude of the excess current.  Since in the
present case $I_2 \sim D$, one would expect that also $I_{exc} \sim
D$. However, it can be shown that the single-particle current gives a
large negative contribution to the excess current: this cancels the
enhanced two-particle current and leaves an excess current which for
all orientations is proportional to $D^2$ in the low-transparency
limit, just like in the low-temperature superconductors. This implies
that the excess current will not be a true measure of the importance
of the Andreev reflection process as soon as there are resonances
(like the MGS) in the junction.

\subsection{Experiments}

\begin{figure}
\centerline{\includegraphics[width=10cm]{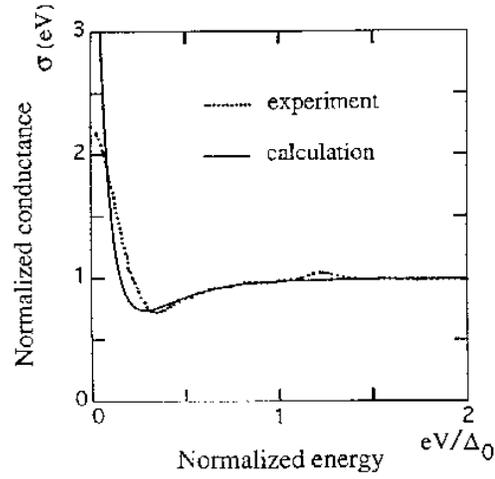}}
\caption{Conductance spectra, normalized by the normal background 
  conductance, obtained on YBCO at $4.2$ K by scanning tunneling
  microscopy fitted by the scattering theory of MGS, assuming the
  $\alpha=\pi/4$ orientation. The parameters of the theory are the
  maximum $d$-wave gap $\Delta_0=17.8$ meV, the normalized barrier
  height $Z=2mH/\hbar^2k_F=3$ (H is the height of the barrier
  potential), and tunnel cone parameter $\beta=10$. From
  \citeasnoun{Kash95}, copyright (1995) by the American Physical
  Society.}\label{Kash_fig1}
\end{figure}

Experimentally, the zero-bias conductance peak (ZBCP) has been studied
for many years: it was observed by \citename{Geerk} already in
\citeyear*{Geerk}.  In the early years, when the $d$-wave symmetry of
the order parameter had not been established, the peak was explained
in terms of a spin-flip scattering model developed by
\citeasnoun{Appelbaum} and \citeasnoun{Anderson}. Following the paper
by \citeasnoun{Hu}, attention was drawn to the MGS scenario. One of
the first good fits between experiments and the $d$-wave theory (MGS,
\fref{Kash_fig1}) was presented by \citeasnoun{Kash95}. Today, a
number of experiments have given quite convincing evidence that the
MGS picture offers the best explanation.  Here we shall discuss some
experimental findings regarding the influence of (1) magnetic field
and (2) orientation and disorder on the MGS and ZBCP. Yet another
class of experiments have been performed recently, namely STM
measurements of MGS bound to point impurities. These experiments will
be discussed in \sref{imp_subsect} below.

\subsubsection{Magnetic field}

\begin{figure}
\centerline{\includegraphics[width=8cm,angle=-90]{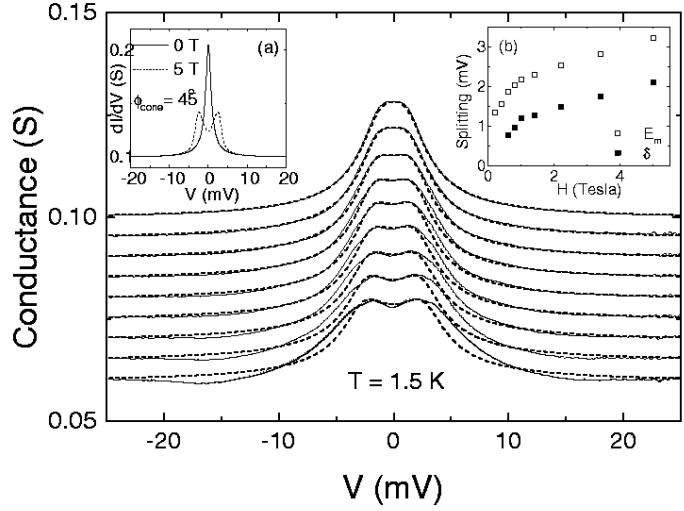}}
\caption{Demonstration of the doppler shift of MGS in the presence of 
  a magnetic field $H=\{0.2,0.4,0.6,0.8,1.0,2.2,3.4,5.0\}T$ in YBCO/Pb
  junctions, where the curve at the top is for the lowest field. The
  spectra are normalized by the high-voltage conductance at $H=0.2T$
  and fitted by a Lorentzian centered at the bound states energy
  $E_B=v_Fp_s\cos\theta$ averaged over angles $\theta$ with a tunnel
  cone $D(\theta)=1$ for $|\theta|<\theta_c=15^o$ and $D(\theta)=0$
  otherwise. The fit parameters are the Lorentzian width
  $\Gamma=[\Delta_0/\Delta(\theta)]\times 0.8$ meV and $v_Fp_s$. In
  inset (a) theoretical results are shown for a larger tunneling cone
  for comparison. In inset (b) the maximum Doppler shift
  $E_m=v_Fp_s\cos(\pi/2-\theta_c)$ and the ZBCP splitting $\delta$ are
  both shown as a function of applied field $H$. From
  \citeasnoun{Aprili}, copyright (1999) by the American Physical
  Society.}\label{Aprili_fig}
\end{figure}

The Doppler shift effect mentioned in \sref{surf_section} was
confirmed experimentally by \citeasnoun{Cov} and may be regarded as
the first successful attempt to seriously discriminate between the MGS
and the spin flip scattering scenarios for the ZBCP.  The effect was
further studied by \citeasnoun{Krupke} and \citeasnoun{Aprili}, see
\fref{Aprili_fig}. The ZBCP was observed to split linearly with $H$ up
to $H_{c1}$, where vortices enter the superconductor and the shift
saturates \cite{FRS}. It was pointed out that the spin-flip model
could be ruled out since the Land\'{e} $g$-factor involved in the
Zeeman type splitting in this case had to be very large in order to
fit the experiment.

In the zero-field limit $H\rightarrow 0$, splitting of the ZBCP was
still observed by \citeasnoun{Cov} and \citeasnoun{Krupke} and also by
\citeasnoun{Kash_JPCS97} and \citeasnoun{Lesueur_JLP99}. This was
interpreted as a sign of spontaneous time-reversal symmetry breaking
at the surface and evidence for a local presence of a subdominant
$s$-wave component, c.f. \sref{surf_section} above. However, a
spontaneous (zero-field) splitting of the ZBCP is not always seen
\cite{Aprili,Yeh}. Moreover, there are other possible explanations for
the splitting than the one involving a subdominant $s$-wave component
of the order parameter; see the discussion of split MGS in
\sref{MGSsplit}.  Also, if for any reason particle-hole symmetry is
broken in the superconductor, the MGS may also be split. For the
identification of the mechanism responsible for the spontaneous
splitting of the ZBCP the same arguments as for the identification of
the origin of the ZBCP are relevant: one needs to measure the response
to some controllable perturbation; e.g. the magnetic field response of
the ZBCP discussed above confirmed that MGS are responsible for the
ZBCP. Since such experiments are lacking for the spontaneously split
ZBCP, we feel that at the present time the existence of a TRSB $d+is$
surface state remains controversial.

\subsubsection{Orientation}

\begin{figure}
\centerline{\includegraphics[width=10cm]{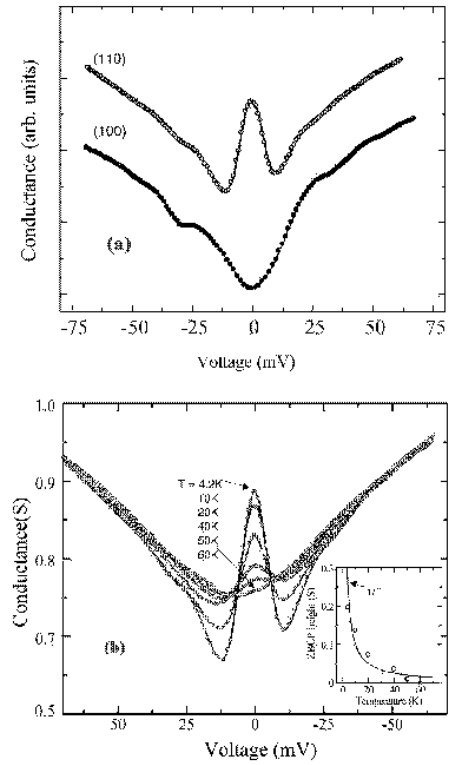}}
\caption{(a) Conductance of YBCO/Ag ramp-edge junctions showing the
  orientation dependence of the zero-bias conductance peak as
  predicted by the MGS scenario: for $\alpha=0$ the peak is absent
  [the (100) curve], while it is present for the $\alpha=\pi/4$
  orientation [the (110) curve]. (b) Temperature dependence of the
  conductance peak [the (110) curve in (a)]. The peak height scales as
  $1/T$, as shown in the inset, while the width is temperature
  independent. From \citeasnoun{WanWang}, copyright (1999) by the
  American Physical Society.}\label{Iguchi_fig}
\end{figure}

As explained above, if we probe an $\alpha=\pi/4$ surface (i.e. [110]
surface) a large ZBCP should be present, while if we probe an
$\alpha=0$ surface (i.e. [100] or [010] surfaces) the ZBCP should be
absent. In some well controlled measurements
\cite{Alff_Rapid,WYGS,WanWang,Iguchi2000} it has been verified that
the ZBCP has this orientation dependence (see \fref{Iguchi_fig}).
However, because of junction inhomogeneities like faceting \cite{FRS},
the local orientation at the junction may not be well defined. For
example the planar junction experiments performed by \citeasnoun{Cov}
showed equal weight of the ZBCP for all orientations of the
electrodes.

\citeasnoun{ACPNG} studied broadening effects on the ZBCP due to
impurities introduced into the junction via ion bombardment.  The
result suggests that disorder introduces decoherence which removes the
MGS resonance. \citeasnoun{ACPNG} argue that if the spin-flip
mechanism were responsible for the ZBCP, the disorder should not have
such a large impact, indicating that the MGS are responsible for the
ZBCP.

\Fref{Iguchi_fig} shows that the peak height decreases as $1/T$ while
the width seems to remain roughly constant, independent of
temperature.  This information is obtained by assuming that the peak
is positioned on a background in the form of a temperature-dependent
gap. In that case, the FWHM is roughly the distance between the two
"fix points" in \fref{Iguchi_fig}. This behavior implies that the
density of MGS (the area of the peak) decreases with increasing
temperature, which can not be explained by thermal smearing. The
question of temperature-independent width has been addressed by
\citeasnoun{Walker} in terms of the effect of rough interfaces and
surface umklapp scattering.

\subsubsection{Impurity induced MGS}\label{imp_subsect}

\begin{figure}
\centerline{\includegraphics[width=10cm]{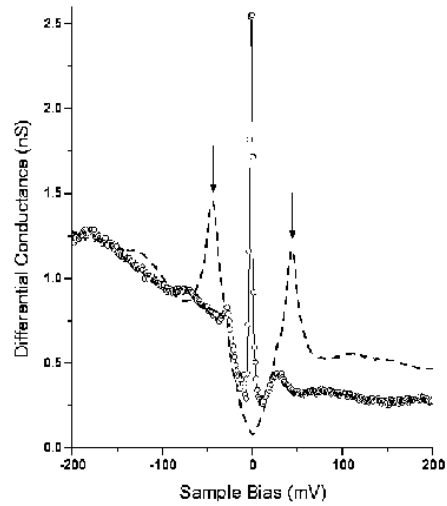}}
\caption{Conductance spectra of Zn-doped BSCCO obtained by scanning
  tunneling microscopy in the $c$-axis direction at $4.2$ K. The
  spectrum taken at a Zn impurity site contains a huge zero-bias
  anomaly (open circles), while the spectrum taken at a clean region
  is of the usual $d$-wave form with peaks at the gap voltages
  (arrows).  Reprinted by permission from Nature [\citeasnoun{Pan}]
  copyright (2000) Macmillan Magazines Ltd.}\label{Pan_fig}
\end{figure}

Bound states induced by impurities have been found experimentally by
different groups \cite{Yazdani,Hudson,Pan}. In the most recent
experiment by \citeasnoun{Pan} the LDOS was measured by STM in the
vicinity of a single impurity (Zn) replacing a Cu atom in the
superconducting plane of BSCCO (\fref{Pan_fig}).  It was found that
the LDOS is enhanced at low energy and it has an anisotropic form in
real space. This observation is in qualitatively good agreement with
theoretical predictions; see
\cite{BalSalRos,BalSal,SalBalSca,FlaBye,Tsu1,Tsu2,Zhu,Zhu_prb00,Maki}
and references therein. It was predicted that a point-like impurity in
a d-wave superconductor is able to form a resonant state within the
superconducting gap. If global particle-hole symmetry is present, then
in the unitary limit (strong impurity potential), the energy of the
resonance approaches the Fermi level, and the resonance width tends to
zero, the MGS becoming a true bound state.

\begin{figure}
\centerline{\includegraphics[width=8cm]{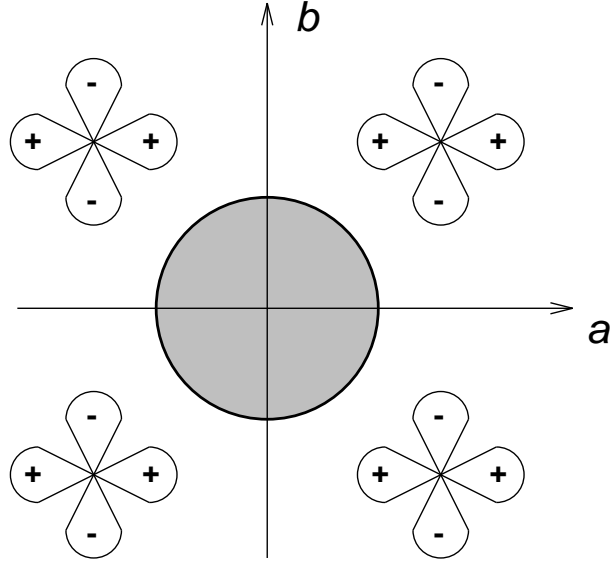}}
\caption{The {\it local} orientation at a circular impenetrable extended
  impurity in a two-dimensional $d$-wave superconductor favors the
  formation of MGS. In real space the wavefunction leaks out along the
  gap nodes, shown in the figure, giving a cross shaped density of
  states similar to what was observed in the experiments by
  \citeasnoun{Pan}.}\label{impurity_fig}
\end{figure}

We can understand this effect by considering a defect with a hard
specular boundary the size of which is much larger than the coherence
length, as discussed by \citeasnoun{MatKanNis}. In this case, the
defect surface can be locally approximated by a planar surface, and
our arguments in favor of the presence of the MGS holds. Since
different parts of the defect boundary have different orientations
with respect to the crystallographic axes and continuously change from
$\alpha=0$ to $\alpha=2\pi$, there are different amounts of
trajectories containing local MGS. Consider for example a circular
defect, as illustrated in \fref{impurity_fig}. Then for the (110) and
(-110) directions the local angle is $\alpha=\pi/4(1+2n)$ which
corresponds to a maximum amount of MGS trajectories, while for (100)
and (010) orientations, the local angle is $\alpha=n\pi/2$ which
corresponds to a minimum amount of MGS trajectories. Hence, the LDOS
at low energy \cite{MatKanNis} is cross shaped in space with tails
along the nodes.  The density of states decreases far from the defect
because the MGS are localized near the scatterer over a distance of
the order of the coherence length.

In fact, the above arguments rely entirely on quasiclassical dynamics
of superconducting quasiparticles, and can therefore be extended (see
\ref{App_MGS}) to defects smaller than the coherence length but larger
than the Fermi wave length \cite{Adagideli}. However, in high-$T_c$
materials this difference is not that important since the coherence
length is rather small.

At the time of writing, an issue under debate is whether the impurity
MGS are localized or extended, see e.g. \cite{ZhuLocalization}. In a
large crystal there are many impurities, or inhomogeneities, and hence
many MGS, which in principle can form an impurity band similar to
impurity bands in semiconductors.  In simplified terms, the MGS
wavefunction can leak out along the gap nodes (the cross-shaped DOS
described above); if wavefunction overlaps between neighboring
impurity states are significant, an impurity band may form. Also
experimentally, the issue is controversial.  Thermal conductivity
measurements \cite{Taillefer_PRL_97,Behnia_JLTP_99} on Y123 and BSCCO
indicate the existence of extended quasiparticle states, while other
investigations \cite{Hussey} on under-doped Y124 indicate the
opposite.

\section{Voltage-biased d/d junctions}

In this Section we study the ac Josephson effect in superconducting
junctions. We begin by outlining the microscopic theory of charge
transport in voltage biased Josephson junctions. Then we discuss how
the MGS resonance appears in this case. In particular, we show that
depending on the lifetime ratio $q$ introduced in the previous
section, the MGS resonance appears either in the pair current at the
gap voltage $eV=\Delta$ (for small intrinsic broadening, $q\ll 1$), or
in the single particle current both near zero voltage and at the gap
voltage (for large intrinsic broadening, $q\gg 1$). In the end we
discuss some experiments.

\subsection{Theoretical formulation}

Generalization of the previous theory of transport through $N/d$
junctions to $d/d$ junctions [beyond the tunnel model;
equation~\eref{tunnel_formula}] is non-trivial due to the Josephson
effect. In the presence of a bias voltage, the phase difference over
the Josephson junction becomes time-dependent,
$\partial_t\phi=2eV/\hbar$ \cite{Jos}, and a quasiparticle incident on
the junction will see the junction as a scatterer whose properties
periodically change in time with the Josephson frequency $\omega_J=
2eV/\hbar$.  In the limit of small applied voltage, $eV\ll\Delta$, the
time-dependent current can be understood via adiabatic arguments, in
terms of a time-dependent phase, letting $\phi\rightarrow \phi(t)
=2eVt/\hbar$ in the equation for the dc Josephson current
\cite{AveBar,BSBW}. In the general case, a full quantum mechanical
consideration is necessary for inelastic scattering by the
time-dependent scatterer, taking into account transitions between the
incident quasiparticle energy $E$ and an infinite set of sideband
energies $E_n=E+neV$ ($n$ are integers). The time-dependent current
can then be decomposed into a spectrum of frequency components
\begin{equation}\label{ac_current}
I(V,t) = \sum_{m=-\infty}^{\infty} I_m(V) e^{im\omega_J t},
\hspace{0.5cm} I_{-m}(V) = I_m^*(V),
\end{equation}
where the $m=0$ component is the time averaged current and the $m\neq
0$ components correspond to the ac oscillations with basic frequency
$\omega_J=2eV/\hbar$. Interestingly, the very early observation by
\citeasnoun{Esteve_87} of the ac Josephson effect was taken as one of
the first proofs for the existence of Cooper pairs in the high-$T_c$
superconductors.

Within the adiabatic picture for the $d$-wave junctions, the
$\pi$-periodic current-phase relation in the $d_0/d_{\pi/4}$ junction
(see \fref{d0d45_fig}) will result in an ac Josephson oscillation with
twice the usual frequency, $2\omega_J$. This was predicted by
\citename{Yip1} \citeyear{Yip1,Yip2} and \citeasnoun{Zago} using the
adiabaticity arguments, and it was recently confirmed by a full
calculation of equation~\eref{ac_current} by \citeasnoun{LJHW}.
Possibly this effect was observed in experiments by
\citeasnoun{Early_APL93}, although the authors presented an
alternative explanation in terms of parallel junctions in the grain
boundary.

The time-independent ($m=0$) current, from now on called the dc
current, has contributions from all sidebands, $-\infty < E_n <
\infty$, and at low temperature it has the form
\begin{equation}\label{n_currents}
I_{dc}(V) = \sum_{n=1}^{\infty} I_n^{dc}(V).
\end{equation}
Remarkably, this expansion for the dc current has two different
physical interpretations, which in fact are equivalent, namely
multiparticle tunneling (MPT) and multiple Andreev reflection (MAR). A
consistent theory unifying the concepts of MPT and MAR has been
presented by \citeasnoun{BSW} and \citeasnoun{Sh}.

The first interpretation, multiparticle tunneling (MPT), is
particularly useful in the tunnel limit, $D\ll1$. It was originally
suggested by \citeasnoun{SchWil} using perturbative tunnel model
calculations, and it was improved by \citeasnoun{Cuevas}.  According
to this interpretation the $n=1$ current, $I_1^{dc}(V)$, is due to
single-particle tunneling processes, and it corresponds to the
quasiparticle current in a standard tunnel model calculation. The
$n=2$ current, $I_2^{dc}(V)$, is due to two-particle tunneling
processes which have smaller probability, $\sim D^2$, and so on. In
s-wave junctions, energy conservation prohibits the $n$-particle
tunneling process for $neV<2\Delta$. Thus, each $n$-particle current
has a threshold at voltage $eV_n=2\Delta/n$ and the magnitude of this
current is proportional to $D^n$. However, resonances due to the
presence of surface states or resonances in the barrier may lower the
exponent and enhance the intensity of the process \cite{Joh,MAR_res}.

\begin{figure}
\centerline{\includegraphics[width=10cm]{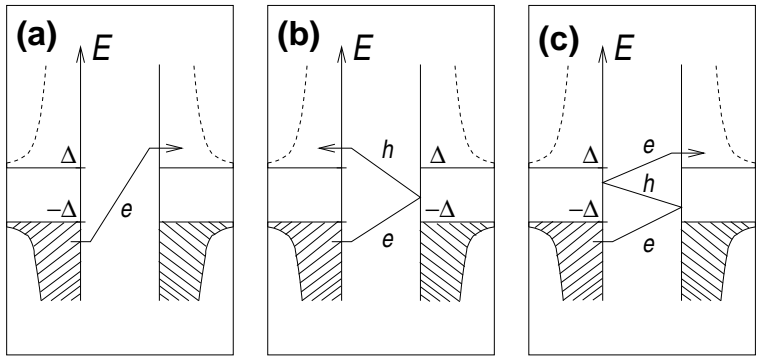}}
\caption{Illustration of the (a) single-particle, (b) two-particle,
  and (c) three-particle processes in the SIS junction. At zero
  temperature allowed processes connect occupied states below the gap
  in the left superconductor with unoccupied states above the gap in
  the superconductors. We have aligned the Fermi energies of the two
  sides by a gauge transformation in the right superconductor;
  consequently, quasiparticles are accelerated by $eV$ at each passing
  through the junction region.}\label{MAR_proc}
\end{figure}

The second interpretation, more useful for transparent junctions, is
based on the scattering theory and involves multiple Andreev
reflections (MAR) \cite{KBT,Arnold,BSW,AveBar}. According to this
interpretation, the current $I_1^{dc}(V)$ is due to direct scattering
(acceleration) across the energy gap, see \fref{MAR_proc}(a), while
the current $I_2^{dc}(V)$ is due to the Andreev reflection processes,
in the same way as in the NS junction, equation~\eref{BTK}, depicted
in \fref{MAR_proc}(b).  For smaller voltage, the Andreev reflected
hole may be Andreev reflected once more when reaching the left
superconductor, this time as an electron [see Fig.~\ref{MAR_proc}(c)],
giving the current $I_3^{dc}(V)$, and so on. The multiple Andreev
reflections form, in principle, an infinite series of currents,
equation~\eref{n_currents}. The magnitude of each current is
proportional to $D^n$ (in the absence of resonances), since each
process contains $n$ coherent passages through the barrier.

The equivalence of these two interpretations $-$ MPT and MAR $-$ is
due to the fact that at every Andreev reflection one pair of electrons
is transmitted through the SN interface, and therefore the MAR process
with $n$ Andreev reflections transfers a charge $(n+1)e$,
corresponding to the $n+1$-particle current \cite{Arnold,Sh}.

In $s$-wave junctions, the onsets of the $n$-particle currents at
$eV=2\Delta/n$, together with peaks at the same voltages in the higher
order currents due to the DOS peaks near the gap edges, is the
explanation for the so-called subharmonic gap structure (SGS), see
\cite{Ludoph} and references therein.

\begin{figure}
\centerline{\includegraphics[width=8cm]{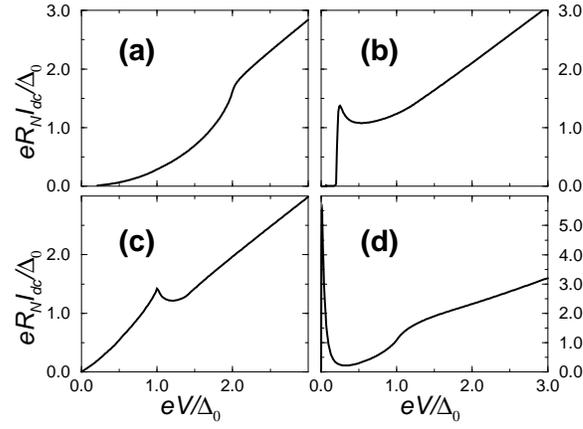}}
\caption{The total dc current in junctions with orientations (a)
  $d_0/d_{0}$, (b) $s/d_{\pi/4}$ where $\Delta_s=0.2\Delta_0$
  ($\Delta_s$ is the $s$-save superconducing gap and $\Delta_0$ is the
  maximum $d$-wave gap), (c) $d_{0}/d_{\pi/4}$, and (d)
  $d_{\pi/4}/d_{\pi/4}$.  The transparency of the tunnel barrier is
  $<D>=0.026$ in all cases and we assume zero
  temperature.}\label{dd_IVtot}
\end{figure}

In d-wave superconductors, the sub-gap structure
\cite{Hurd,BarSvi,HLJW,LJSWH,Samanta,Yoshida,LJW} is very different
for two reasons: (1) the multiparticle current thresholds are washed
out due to the gap nodes, (2) the presence of MGS changes the
resonances in energy space. For the $d_0/d_0$ orientation (no MGS in
the junction), the single-particle current dominates the $I-V$
characteristics because of the gap nodes. Higher order processes
($n\geq 2$) are very small corrections and can be neglected; thus,
sub-harmonic gap structure at $eV_n=2\Delta/n$ is suppressed in the
I-V characteristics in $d$-wave junctions, see Fig.~\ref{dd_IVtot}(a).
When MGS are present in the junction, new resonances appear. In
particular, the two-particle process becomes resonant (see below) and
must be taken into account.  Although the MGS may enhance also higher
order processes than $n=2$, these resonances are always very weak
compared to the two-particle process. For these reasons we will in the
following only discuss the single particle and two particle currents.

We start our discussion of the $I-V$ characteristics of junctions with
MGS by presenting in \fref{dd_IVtot}(b)-(d) results of full numerical
calculations of equation~\eref{n_currents} for junctions where the
right electrode has $d$-wave symmetry with orientation
$\alpha_R=\pi/4$ and the left counter electrode has $s$-wave symmetry
($s/d_{\pi/4}$ junction), $d$-wave symmetry with $\alpha_L=0$
orientation ($d_0/d_{\pi/4}$ junction), and $d$-wave symmetry with
$\alpha_L=\pi/4$ orientation ($d_{\pi/4}/d_{\pi/4}$ junction),
respectively. In the last junction case we introduced a small
imaginary part (to model broadening) to the energy in order to reveal
the current peak near zero bias. If the MGS are strongly coupled to
the reservoirs (large intrinsic broadening, $q\gg 1$), the
current-voltage characteristics in \fref{dd_IVtot} are dominated by
single particle tunneling, and they can be qualitatively understood by
applying tunnel model arguments. Indeed, when the right superconductor
is a $d$-wave superconductor with the $\pi/4$ orientation, the local
density of states is given by equation~\eref{LDOS_MGS}; the MGS delta
function $\delta(E)$ then produces a term \cite{Samanta}
\begin{equation}\label{SS_tunneling_MGS}
I_1^{dc}(V)\propto\left<D(\theta)
\left[n_F(-eV)-n_F(0)\right]
\rho_L(\theta,-eV)\right>.
\end{equation}
Within this model, we understand figures \ref{dd_IVtot}(b)-(d) as
simply the $s$-wave DOS, the angle integrated $d$-wave DOS, and
finally the midgap peak. Note that the {\em current itself} rather
than the conductance is proportional to the LDOS $\rho_L$ of the left
superconductor. For voltages $eV>\Delta$ terms not included in
equation~\eref{SS_tunneling_MGS} become important and for
$eV\gg\Delta$ the current is proportional to the applied voltage.

\begin{figure}
\centerline{\includegraphics[width=8cm]{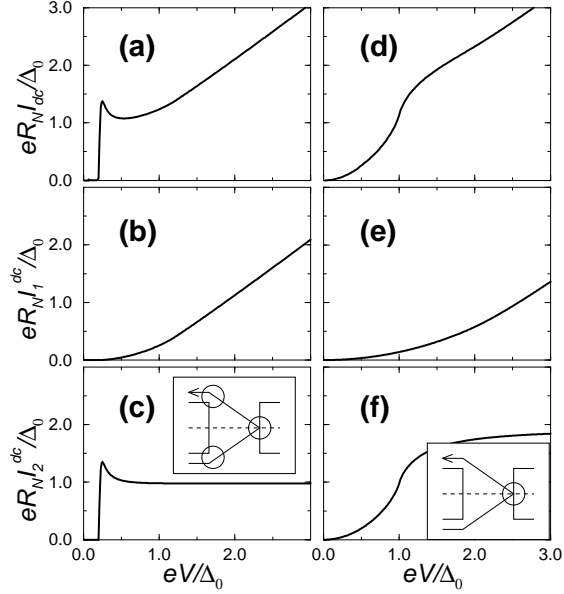}}
\caption{(a)-(c) The total current [the same curve as in
  \fref{dd_IVtot}(b)], the one-particle current, and the two-particle
  current in the $s/d_{\pi/4}$ junction with $\Delta_s=0.2\Delta_0$
  where $\Delta_s$ is the $s$-save superconducing gap and $\Delta_0$
  is the maximum $d$-wave gap. The peak at the gap voltage is due to
  the MGS resonance on the right side overlapping with the gap-edge
  resonances at the entrance and exit points on the $s$-wave
  superconductors side, see inset in (c) where the circles indicate
  presence of resonances.  Equivalently, we can say that the large BCS
  gap-edge density of states pushes more current through the MGS,
  compared to the NS case. (d)-(f) The total current, the one-particle
  current, and the two-particle current in the $d_{\pi/4}/d_{\pi/4}$
  junction. The onset at the gap voltage is due to the {\it bare} MGS
  resonance, see inset in (f) where the circle indicate presence of
  resonance. The transparency of the tunnel barrier is $<D>=0.026$ in
  all cases and we assume zero temperature.}\label{dd_IV}
\end{figure}

When the MGS are intrinsically very narrow ($q\ll 1$), the resonance
will not appear in the single-particle current, instead the resonance
appears in the two-particle current (compare the discussion of the
$N/d$ junction in \sref{Nd_sec}). We exemplify this with the
$s/d_{\pi/4}$ and $d_{\pi/4}/d_{\pi/4}$ junction cases by plotting the
single- and two-particle currents separately along with the total
currents in \fref{dd_IV}. In the $s/d_{\pi/4}$ junction case, the I-V
relations (total current) for the $q\gg 1$ and $q\ll 1$ cases are
similar, although in the $q\ll 1$ limit the peak at the $s$-wave gap
voltage is due to the two-particle current; the resonant process is
shown in the inset. This situation is equivalent to the one in the N/d
case discussed in \sref{Nd_sec}. However, in the $d_{\pi/4}/d_{\pi/4}$
junctions, which have MGS at both sides of the junction barrier, the
situation is different. The process of MGS-to-MGS tunneling is absent
in the small $q$ limit. Thus there is no current peak at small voltage
in this case [compare \fref{dd_IVtot}(d) and \fref{dd_IV}(d)]; instead
the resonance shows up in the two-particle current at the gap-voltage
$eV=\Delta$, see \fref{dd_IV}(f); the resonant process is shown in the
inset. Thus, for this orientation it should be possible experimentally
to distinguish the two cases, large or small $q$, by observing where
the resonance appears: either near zero voltage ($q\gg 1$) or at the
gap voltage ($q\ll 1$), respectively.

\subsection{Comments on experimental results}
A large number of current-voltage characteristics (IVC) measurements
have been performed with grain boundary and ramp-edge junctions
\cite{DelinKleinsasser}. In this Review we will not make any serious
attempt to analyze these experiments beyond what is stated in the
original papers. Instead we will make some comments on a few specific
issues.

Measurements of IVC of $d/d$ junctions are difficult to analyze.
Besides complications with the samples (see the discussion of
measurements of the critical Josephson current in
Section~\ref{Ic_exp}), the presence of the Josephson effect makes the
sub-gap voltage region hard to interprete.

Because of the Josephson current, an essential issue concerns how to
measure IVC.  In order to measure quasi-particle current and
conductivity, and to observe the MGS zero-bias peaks, one must work
with voltage bias, which by definition avoids the problem with the dc
Josephson current. However, this is hard to do experimentally, because
of the low junction impedance; in practice, current bias is therefore
usually employed.

Alff et al. \citeyear{Alff_Eur,Alff_prb} have recently reported IVC
measurements on symmetric $d_{\alpha}/d_{-\alpha}$ HTS grain boundary
junctions with large misorientation angles, in the range $30-45^0$.
These junctions show very small or vanishing Josephson current, and
the conductance $dI/dV$ vs $V$ shows a zero-energy structure quite
similar to NIS junctions (see \fref{Iguchi_fig}), with similar
temperature and magnetic field dependences.  Alff et al.
\citeyear{Alff_Eur,Alff_prb} conclude that they in fact observe
NIS-like behavior, although they interpret the zero-energy structure
in the $dI/dV$ vs $V$ characteristic as a ZBCP resulting from an SIS
convolution of density of states with strongly broadened MGS.

The similarity of the ZBCPs of the NIS and SIS junctions is striking,
and we would like to consider the following tentative explanation. For
an SIS junction to behave like an NIS junction, it is necessary that
there are very strong relaxation effects in the barrier region, the
barrier acting as a reservoir, which results in decoupling of the two
SI and IS interfaces. The SI/IS junctions will then be in series and
each will be biased by $eV/2$; as a result, there will be a zero-bias
onset of the current (without peak) as in the NIS junction, but the
maximum gap should occur at $2\Delta$. The average gap voltage $V_g$
could be well below this value.  Whether this agrees with the
experimental results remains an open question since there is no
independent measurement of the gap in the experiment.

Alff et al. \citeyear{Alff_Eur,Alff_prb} rather consider their results
to signify that the junction is in the {\em tunneling} regime, SIS,
albeit tunneling is considered to proceed through resonant states in
the barrier.  Let us then assume that we can describe the junction as
a "tunnel" junction S"I"S, where the "I"-region cannot be regarded a
reservoir.  The Josephson current could still be suppressed for
several reasons: the junction could be long in comparison with the
coherence length, or suppressed by inelastic interactions, or there
could be fluctuations and noise causing the two sides of the junction
to become incoherent. In a symmetric situation with narrow MGS on both
sides of the junction, as described before, the MGS DOS peaks will
then be convoluted, leading to a current peak, i.e. to a peak in the
IVC and not in the conductance.  In this view, the conductance
zero-energy anomaly in the $d_{\alpha}/d_{-\alpha}$ junction describes
the {\em voltage derivative} of the MGS DOS, and should not be
directly compared with the MGS DOS of the NIS junction.

Note that if the MGS is very broad, say $\Gamma\sim 0.3 \Delta_0$, the
MGS peak in the calculated IVC will merge into the rising background,
and the result may {\em appear similar} to the NIS case; nevertheless,
in our opinion, one should still look for the MGS DOS in the IVC, and
not in the conductance. However, with such extreme broadening the MGS
may act as normal reservoirs in the gap, and the whole situation may
have to be reconsidered.

In conclusion, we believe that further progress depends on close
interaction between experiment and theory to critically examine models
and results in order to produce internally consistent pictures and new
critical experimental tests.

\section{Other experiments}

\subsection{$c$-axis tunneling}

For a long time, the presence of a Josephson current in experiments on
$c$-axis tunnel junctions between YBCO and Pb has been regarded as
rather mysterious, because if the order parameter of YBCO has pure
$d$-wave symmetry, the first order Josephson coupling should vanish,
only leaving a small residual second order current (see e.g.
references~\cite{Sun_PRL94,Kleiner_PRL96,Lesueur_PRB97}. This
conclusion however relies on the assumption that the c-axis current
injection, from e.g. an STM tip, results in a spherically symmetric
ab-plane injection current. Only under such circumstances is the
overlap between the $s$-wave and the $d$-wave order parameters zero.
Considering spherical injection, the orthorhombic structure of YBCO
indicates that there may be, in addition to the $d$-wave order
parameter, a small $s$-wave component present which could be
responsible for the first order Josephson coupling observed in the
experiments. However, twinning spoils this explanation since
neighboring twins should have different signs of the $s$-wave
component, and the average Josephson current should be close to zero.
\citeasnoun{Kouz_PRL97} studied tunneling along the $c$-axis from Pb
into YBCO containing one single twin boundary. It was shown that
indeed a $d\pm s$ order parameter is realized in YBCO and that the
sign between the $d$-wave and $s$-wave components is opposite on the
two opposite sides of the twin boundary.  However, also in experiments
on junctions between Pb and highly twinned YBCO \cite{Kleiner_PRL96},
a first order Josephson coupling was found.  Furthermore, in
experiments on $c$-axis BSCCO-Pb junctions \cite{Mossler_PRB99} a
finite first-order Josephson coupling was observed and interpreted in
terms of a small $s$-wave component also in BSCCO. Recently an
interesting possibility was brought forward by \citeasnoun{Rae_PRL00},
who pointed out that Pb is a rather anisotropic $s$-wave
superconductor and showed how this asymmetry on the $s$-wave
superconductor side can explain at least the experiments on BSCCO. To
test this idea, a thorough investigation of the Josephson coupling
between isotropic $s$-wave superconductors and BSCCO should be
performed.

\subsection{Electron-doped cuprates}

Very recently, \citeasnoun{tricry_NCCO} found the half-flux quantum
also in tricrystal rings of the electron-doped cuprates NCCO and PCCO.
This provides very strong evidence for $d$-wave superconductivity in
both these materials. This conclusion is supported by the power-law
dependence of the penetration depth found by \citeasnoun{Prozorov}.
However, in transport experiments
\cite{Ekin,Alff_Eur,KashNCCO,Klee,AlffNCCO} ZBCPs were never found.
Instead, more or less V-shaped spectra were obtained.  This absence of
ZBPC was taken as evidence for an anisotropic $s$-wave order parameter
in NCCO, but now turns out to be in disagreement with the very recent
tricrystal ring experiments \cite{tricry_NCCO}.  Therefore, at this
point, a reexamination of the transport properties of the
electron-doped materials seems necessary in order to check the
presence of MGS also in these materials. We would like to point out
that the experimental results obtained by Alff can be understood
within the MGS scenario if the MGS in these materials are
intrinsically very narrow [small $q$, equation~\eref{tau_quota}].
However, also in the normal-metal-superconductor (N/d) experiments on
NCCO by \citeasnoun{KashNCCO} the ZBCP was absent, which can not be
explained by a narrow MGS only.

\section{Summary and concluding remarks}

Midgap states (MGS) are consequences of the angle-dependent sign
change of the order parameter and provide direct evidence for $d$-wave
symmetry. Transport measurements which detect the MGS resonance at the
Fermi energy for particular orientations of the plane of the junction
with respect to the crystallographic axes will provide crucial
evidence for $d$-wave superconductivity.  The MGS phenomenon is part
of a general problem of superconducting surface and interface states
and their relation to the Andreev states, which determine the
transport properties of superconducting junctions.

Surface midgap states (MGS) in d-wave superconductors are robust
features which result in a large number of observable effects: large
Josephson current, low-temperature anomaly of the critical current,
$\pi$-junction behavior, $0\rightarrow \pi$ junction crossover with
temperature, zero-bias conductance peaks, paramagnetic currents, time
reversal symmetry breaking and spontaneous interface currents, and
resonance features in subgap currents.

Some of these consequences of MGS have been verified experimentally
(zero-bias conductance peaks, paramagnetic currents and
impurity-induced states), while some remain controversial
(time-reversal symmetry breaking and spontaneous surface/interface
currents), or so far unobserved (large Josephson current and the $1/T$
low-temperature anomaly of the critical current in low-transparency
junctions).

A number of problems need to be clarified, e.g. the character and
strength of relaxation processes and the width of the MGS.  In
particular, does absence of MGS imply absence of $d$-wave (and
presence of $s$-wave), or does it imply $d$-wave and strong
decoherence?

The predicted low-temperature anomaly associated with MGS, $I_c
\propto 1/T$, remains to be confirmed experimentally. The lack of
experimental evidence may be due to faceting of the grain-boundary
\cite{Mannhart_PRB96}: different parts of the junction may have local
orientations supporting large MGS currents, but with opposite signs
for different facets.  Consequently, the low-temperature anomaly
associated with the MGS may be averaged out for a macroscopic
junction. The obvious remedy is to work with junctions with interfaces
with well-defined orientation.

The true nature of HTS grain-boundary (GB) and step-edge Josephson
junctions remains, in our opinion, a mystery. It is difficult to get a
consistent picture. In recent work on low-impedance step-edge
junctions, a reasonable description of the Josephson current could be
obtained assuming high transparency, while at the same time the low
junction resistivity suggested that the transport is through
high-transparency spots, point contacts or pinholes, in the interface
barrier. In GB junctions at moderate misorientation angles a similar
description may be appropriate. At large misorientation angles,
however, the GB junctions has been characterized as an SIS tunneling
junction with resonant states in the barrier. However, it is difficult
to judge whether the experimental results imply coherent tunneling or
incoherent transport through the junction. Here coevolution of
experiment and theory seems to be necessary to obtain deeper
understanding.

Other predictions which need to be verified experimentally include
negative differential conductance in sub-gap structure (needs a
voltage-bias setup), and spontaneous magnetization in time-reversal
symmetry breaking (TRSB) states at $d_0/d_{\pi/4}$ junctions.

In a number of interesting recent experiments, a picture has emerged
that the superconducting order parameter of the cuprates might not
have pure $d_{x^2-y^2}$ symmetry but rather have a complex order
parameter with $d_{x^2-y^2}+id_{xy}$ symmetry, where the $d_{xy}$
component is small compared to the $d_{x^2-y^2}$ component.  STM
studies of vortex cores in BSCCO \cite{Pan_vortex} have revealed
enhanced density of states at energies near $\pm 7$ meV (i.e. well
inside the bulk superconducting gap) with isotropic exponential
dependence on distance from the core center. This might indicate that
a field induced $d_{x^2-y^2}+id_{xy}$ order parameter is realized, see
\cite{Franz,Pan_vortex} and references therein. \citeasnoun{Tafuri}
have recently found spontaneous circulating currents around impurities
in YBCO where the vortices contained fluxes of fractional size, in
line with the idea of a complex order parameter induced near
(magnetic) impurities \cite{Graf_PRB00}.
\citeasnoun{Movshovich_PRL98} found that the thermal conductivity of
nickel-doped BSCCO suddenly dropped at about $200$ mK, which was taken
as evidence for a second superconducting transition to a state with
$d_{x^2-y^2}+id_{xy}$ order. On the other hand, in a very recent
experiment reported by \citeasnoun{Carmi_Nature00}, spontaneous
magnetization of {\em bulk} YBCO was found to appear at the
superconducting transition temperature, meaning that the {\em bulk}
order parameter may have complex $d_{x^2-y^2}+id_{xy}$ symmetry.  At
this point, it seems crucial to collect more evidence for (or against)
the $d_{x^2-y^2}+id_{xy}$ order and establish whether it appears in
the bulk at $T_c$, in magnetic fields, or near imperfections like
magnetic impurities at low temperature.

Finally, the obvious, but perhaps naive, question concerns the
all-important issue of the mechanism for high-$T_c$ superconductivity,
and the true symmetry of the order parameter. Despite the absence of a
definitive theory of the pairing mechanism and symmetry of the order
parameter - a consistent picture of $d$-wave singlet pairing and
Andreev midgap states is slowly emerging, unifying a large body of
experimental results and providing a framework for critical
examination of present and future experiments. This means, perhaps,
that inconsistent results may be discarded and significant new effects
discovered and accepted. If another paradigm shift becomes
unavoidable, e.g. based on a consistent non-Fermi-liquid picture of
the normal state, it is likely that this must incorporate the present
notion of Andreev midgap states.

\ack

We are grateful to L. Alff, M. Fogelstr\"om, R. Gross, I. Iguchi, Z.
Ivanov, Y. Tanaka, A. Yurgens, J.-X. Zhu, and S. \"Ostlund for
valuable discussions. We also thank T. Claesson, C.  C.  Tsuei, and A.
Zagoskin for reading and commenting on the manuscript.  Part of the
studies included in the review was done during the visit of one of us
(VS) to the Center for Advanced Studies in Oslo; VS is grateful to the
administration of the Center and particularly to Yu.  Galperin for
warm hospitality and support.  This work has been supported by grants
from NFR and NUTEK (Sweden) and NEDO (Japan).

\appendix
\section*{Appendix}
\setcounter{section}{1}

In this appendix we provide a simple derivation of the midgap states
(MGS).  In some detail, we first solve the Bogoliubov-de Gennes
equation for bulk $s$-wave and $d$-wave superconductors and then we
discuss the clean NS interface and the concept of Andreev reflection.
The Andreev reflection phase shift used throughout the review is then
obtained. Finally we show how the MGS are formed as a consequence of
the sign difference between the lobes of the $d$-wave gap and why the
formation of MGS is not sensitive to the detailed spatial dependence
of the gap, although the MGS wavefunction itself depends on this
shape.

Some particularly useful references regarding the technique used to
obtain the solutions of the Bogoliubov-de Gennes equation for
different structures can be found in \cite{BTK,Bruder,Bagwell_PRB92}.
For the $s$-wave NS junctions we refer to \citeasnoun{BTK} and for the
$d$-wave case we refer to \citeasnoun{Bruder}.
\citeasnoun{Bagwell_PRB92} derived the Andreev states present in
$s$-wave SIS structures.

\subsection{Plane wave solutions of Bogoliubov-de Gennes equation}

The BdG equations for $s$-wave superconductors are
\begin{equation}
\left\{
\begin{array}{lcl}
H_0 u({\bf r}) + \Delta({\bf r}) v({\bf r})&=&Eu({\bf r})\\
 -H_0v({\bf r}) + \Delta^*({\bf r}) u({\bf r})&=&E v({\bf r}),
\end{array}
\right.\label{BdG}
\end{equation}
where $u$ and $v$ are the electron and hole components of the
wavefunction, $H_0=-\hbar^2\nabla^2/2m-\mu$ in the simplest case, and
$E$ is measured relative to the chemical potential $\mu$.

\Eref{BdG} is easily solved for a bulk superconductor with
$\Delta({\bf r})=\Delta=const$ by making the following ansatz:
\begin{equation}
\Psi_S({\bf r})= \left(
\begin{array}{c}
u\\ v
\end{array}
\right)e^{i{\bf k}\cdot{\bf r}},
\end{equation}
where $u$ and $v$ do not depend on ${\bf r}$. The secular equation
restricts the modulus of the ${\bf k}$-vectors, ${\bf k} = {\bf n}k$,
to
\begin{equation}
k=\sqrt{\frac{2m}{\hbar^2}(\mu\pm\sqrt{E^2-\Delta^2})}\approx
k_F\pm\frac{\xi}{\hbar v_F},
\end{equation}
where $\xi=\sqrt{E^2-\Delta^2}$. In the last step we assumed that the
quasiparticle energy is much smaller than the Fermi energy.  There
will then be two types of wave functions, electron-like and hole-like,
depending of what type of solutions they describe in the normal limit
($\Delta\rightarrow 0$). Labeling these by $k^e$ and $k^h$
respectively, where $k^{e,h}=k_F\pm\sigma\xi/\hbar v_F$
($\sigma=\mbox{sgn}E$), we have
\begin{equation}
\Psi^e_S=\left(
\begin{array}{c}
u_0\\ v_0
\end{array}
\right) e^{ik^e{\bf n}\cdot{\bf r}},\hspace{1cm} \Psi^h_S=\left(
\begin{array}{c}
v_0\\ u_0
\end{array}
\right) e^{ik^h{\bf n}\cdot{\bf r}},
\end{equation}
where $u_0$ and $v_0$ are given by $v_0/u_0=(E-\sigma\xi)/\Delta$
together with the normalization condition $|u_0|^2+|v_0|^2=1$.  Above
we assumed that $|E|>\Delta$ since there cannot be any propagating
solutions for energies below the gap. However, there are also
exponentially decaying or growing solutions which become important for
surfaces. By defining $\xi=i\sigma\sqrt{\Delta^2-E^2}$ for
$|E|<\Delta$, the above expressions describe the exponentially
decaying (in the direction of propagation) solutions for subgap
energies.

For the normal metal, the solutions are obtained from the above by
letting $\Delta\rightarrow 0$. The electron and hole wave functions
can then be written as
\begin{equation}
\Psi_N^e=\left(
\begin{array}{c}
1\\ 0
\end{array}
\right) e^{ik^e_N{\bf n}\cdot{\bf r}},\hspace{1cm} \Psi_N^h=\left(
\begin{array}{c}
0\\ 1
\end{array}
\right) e^{ik^h_N{\bf n}\cdot{\bf r}},
\end{equation}
where $k^{e,h}_N\approx k_F\pm E/\hbar v_F$.

Note that the quasiparticle's direction of propagation is given by the
group velocity $v_g=(1/\hbar)dE/dk$, which for a positive k is in the
positive (negative) direction for electron-like (hole-like)
quasiparticles.

\subsection{d-wave case}

In the $d$-wave case, the Bogoliubov-de Gennes equations are non-local
\begin{equation}
\left\{
\begin{array}{lcl}
H_0 u({\bf r}) + \int d{\bf r}'\Delta({\bf r},{\bf r}') v({\bf r}')
&=& Eu({\bf r})\\
-H_0 v({\bf r}) + \int d{\bf r}' \Delta^*({\bf r},{\bf r}') u({\bf r}')
&=& Ev({\bf r}).
\end{array}
\right.\label{nonlocal_BdG}
\end{equation}
Making the quasiclassical approximation as described in e.g.
reference~\cite{Bruder}, we separate out the fast $1/k_F$ oscillations
from the beginning by making the ansatz
\begin{equation}
\left(
\begin{array}{c}
u({\bf r})\\ v({\bf r})
\end{array}
\right)=\left(
\begin{array}{c}
\tilde{u}({\bf r})\\ \tilde{v}({\bf r})
\end{array}
\right) e^{ik_F{\bf n}\cdot{\bf r}}
\end{equation}
where the spatially slowly varying envelope functions $\tilde{u}({\bf
  r})$ and $\tilde{v}({\bf r})$ satisfy the quasiclassical BdG
equation
\begin{equation}
\left\{
\begin{array}{lcl}
-i\hbar v_F {\bf n}\cdot\nabla \tilde{u}({\bf r})
+\Delta({\bf n, r})\tilde{v}({\bf r})& = & E\tilde{u}({\bf r})\\
i\hbar v_F {\bf n} \cdot\nabla \tilde{v}({\bf r})
+\Delta^*({\bf n, r})\tilde{u}({\bf r}) & = & E\tilde{v}({\bf r})
\end{array}
\right.\label{quasiclassicalBdG}
\end{equation}
where terms of order $1/k_F \xi_0$ and higher have been neglected.
This approximation is valid when the inequality $\Delta\ll E_F$ holds
[see e.g. \cite{Andreev}]. For the spatially independent order
parameter, taking the ${\bf r}$-dependence of $\tilde{u}({\bf r})$ and
$\tilde{v}({\bf r})$ to be $\exp(i\kappa({\bf n}){\bf n}\cdot{\bf
  r})$, the secular equation restricts $\kappa({\bf n})$ to
$\pm\xi({\bf n})/\hbar v_F$, where $\xi({\bf
  n})=\sqrt{E^2-|\Delta({\bf n})|^2}$. In the same way as in the
$s$-wave case, we then find electron-like and hole-like solutions
\begin{equation}\label{dwavefunctions}
\Psi_S^e=\left(
\begin{array}{c}
u_0\\ v_0
\end{array}
\right)e^{ik^e{\bf n}\cdot{\bf r}},\hspace{1cm} \Psi_S^h=\left(
\begin{array}{c}
v_0\\ u_0
\end{array}
\right)e^{ik^h{\bf n}\cdot{\bf r}},
\end{equation}
where $k^{e,h}=k_F\pm\sigma\xi({\bf n})/\hbar v_F$, and $u_0$ and
$v_0$ are given by $v_0/u_0=(E-\sigma\xi({\bf n}))/\Delta({\bf n})$
together with the normalization condition $|u_0|^2+|v_0|^2=1$.

As in the $s$-wave case, for subgap energies we can find exponentially
growing or decaying solutions given by the above expressions if we let
$\xi({\bf n}) = i\sigma\sqrt{|\Delta({\bf n})|^2-E^2}$.

\subsection{Andreev reflection}

On a superconducting surface, electrons are reflected as holes and
vice versa. In this section we will discuss this basic process more
quantitatively by solving the BdG equation for the normal
metal/$s$-wave superconductor interface.  Putting the N/S interface at
$x=0$ and assuming that the gap depends on the space coordinate as
\begin{equation}
\Delta(x)=\left\{
\begin{array}{lc}
0, & x<0\\ \Delta e^{i\chi}, & x>0
\end{array}
\right.\label{delta_stepfunction}
\end{equation}
we are left with the simple problem of matching elementary plane wave
solutions of the BdG equation in each region at the surface. The plane
wave solutions of the BdG equation are derived above in this appendix.
Assuming, for simplicity, propagation along the $x$-axis only, we
write down an ansatz wavefunction describing an electron incident on
the superconductor from the normal metal side
\begin{equation}
\Psi_N= \left(
\begin{array}{c}
1\\ 0
\end{array}
\right)e^{ik^e_N x}+ A\left(
\begin{array}{c}
0\\ 1
\end{array}
\right)e^{ik^h_N x}+ B\left(
\begin{array}{c}
1\\ 0
\end{array}
\right)e^{-ik^e_N x},
\end{equation}
where we have taken into account that the electron may be reflected as
an electron as well as a hole. On the superconductor side we include
transmitted electron-like and hole-like quasiparticles
\begin{equation}
\Psi_S=C\left(
\begin{array}{c}
u_0\\ v_0 e^{-i\chi}
\end{array}
\right) e^{ik^e x}+D\left(
\begin{array}{c}
v_0\\ u_0 e^{-i\chi}
\end{array}
\right) e^{-ik^h x}.
\end{equation}
At the interface (at $x=0$) the wave function continuity gives
\begin{equation}
B=D=0,\; C=1/u_0,\; A=(v_0/u_0)e^{-i\chi}.
\end{equation}
During this process of Andreev reflection, the electron incident from
the normal metal is reflected as a hole with the probability amplitude
$A$. The probability for Andreev reflection is $|A|^2=1$ for energies
within the gap while quickly decreasing outside the gap.  It is then
convenient to write the amplitude of Andreev reflection for
$|E|<\Delta$ on the form
\begin{equation}
A=\sigma e^{-i\gamma} e^{-i\chi}, \;\; \gamma=\arccos \frac{E}{\Delta},
\end{equation}
which defines the Andreev reflection phase shift $\gamma$ used in the
text. Note that also the phase $\chi$ of $\Delta$ is picked up during
Andreev reflection. Charge conservation during Andreev reflection is
preserved by letting a Cooper pair enter the superconductor.
Consequently, a charge $2e$ is transferred from the normal metal to
the superconductor, which explains how subgap current may appear. In
the same way, a hole incident on the superconductor from the normal
metal side may be Andreev reflected as an electron transferring a
charge $2e$ from the superconductor to the normal metal. The phase
shift is in this case $-\gamma+\chi$ as can be checked by repeating
the above calculation for an incident hole, keeping the rest of the
ansatz as it is written.

\subsection{Midgap states}\label{App_MGS}

In this section we discuss quasiparticle reflection from the
insulator/d-wave superconductor interface for energies within the
superconducting gap and the formation of MGS. In the text we discussed
MGS having the structure in \fref{surface_d_fig} in mind. Here we show
an alternative way of finding the MGS, in which it becomes clear that
the normal metal plays no crucial role: the MGS is formed {\em only}
as a consequence of the sign difference between the $d$-wave lobes.
When the size of the normal metal in \fref{surface_d_fig} becomes
vanishingly small, the remaining part of the MGS wavefunction is the
exponential tails in the superconductor. Here we study this
wavefunction first for step-function shape of $\Delta$ and then for
arbitrary shape.

Specular reflection at the surface imposes the following boundary
condition for functions $\Psi_S$ in equation~\eref{dwavefunctions}: the
eigenstate consists of a superposition of states with trajectories
${\bf n}$ and ${\bar{\bf n}}$, and the total wave function is zero at
the surface, i.e.
\begin{equation}\label{BC1}
\Psi_S({\bf n}, 0)= - \Psi_S({\bf{\bar n}},0).
\end{equation}
For spatially constant $\Delta$, the solutions of
equation~\eref{quasiclassicalBdG} with $E=0$ and decaying at infinity
are given by
\begin{eqnarray}\label{states}
\Psi_S({\bf n,r})= {1\choose i\mbox{sgn}\Delta}
e^{ik_F{\bf n}\cdot{\bf r}-|\Delta/\hbar v_{Fx}|x},\nonumber\\
\Psi_S({\bf \bar{n},r})= {1\choose -i\mbox{sgn}\bar\Delta}
e^{ik_F{\bf \bar{n}}\cdot{\bf r} - |\bar\Delta/\hbar v_{Fx}|x}.
\end{eqnarray}
If $\bar\Delta=-\Delta$, these functions have similar vector structure
and they can be matched at the surface, $x=0$, according to
equation~\eref{BC1}, and therefore there are MGS in this case.

\Eref{quasiclassicalBdG} has solutions at $E=0$ also for arbitrary
spatial variation of the gaps $\Delta({\bf r})$ and $\bar\Delta({\bf
  r})$. By introducing the following new variable,
\begin{equation}
\zeta=\int dl|\Delta(l)|/\hbar v_F,
\end{equation}
where $l={\bf n}\cdot {\bf r}$, one can reduce
equation~\eref{quasiclassicalBdG} to:
\begin{equation}
\left\{
\begin{array}{lcl}
\displaystyle -i {d\tilde{u}\over
d\zeta}+\mbox{sgn}(\Delta)\tilde{v}& = & 0\\
\displaystyle i{d\tilde{v}\over d\zeta} +\mbox{sgn}(\Delta)\tilde{u} & = & 0 .
\end{array}
\right.
\end{equation}
These equations have solutions similar to the ones in
equation~\eref{states}, which leads to MGS.

\section*{References}

\Figures

\newpage

\end{document}